\documentclass[letterpaper,10pt]{article}
\usepackage{jheppub}
\usepackage[dvipsnames]{xcolor}
\usepackage{graphbox}

\newcommand{\nn}{\nonumber}
\newcommand{\wc}{\psi}
\newcommand{\swc}{\tilde{\psi}}
\newcommand{\ch}{\text{-chain}}
\newcommand{\G}{\mathcal{G}}
\newcommand{\A}{\mathcal{A}}

\title{Hidden Zeros of the Cosmological Wavefunction}

\author[a,b]{Shounak De,}
\emailAdd{shounak\_de@brown.edu}

\author[a]{Shruti Paranjape,}
\emailAdd{shruti\_paranjape@brown.edu}

\author[a]{Andrzej Pokraka,}
\emailAdd{andrzej\_pokraka@brown.edu}

\author[a,c]{Marcus Spradlin,}
\emailAdd{marcus\_spradlin@brown.edu}

\author[a]{Anastasia Volovich}
\emailAdd{anastasia\_volovich@brown.edu}

\affiliation[a]{Department of Physics,
    Brown University,
    Providence,
    RI 02912,
    USA
}

\affiliation[b]{Kavli Institute for Theoretical Physics,
    University of California,
    Santa Barbara,
    CA 93106,
    USA
}

\affiliation[c]{Brown Theoretical Physics Center,
    Brown University,
    Providence,
    RI 02912,
    USA
}

\abstract{%
Motivated by the recent discovery of hidden zeros in particle and string amplitudes, we characterize zeros of individual graph contributions to the cosmological wavefunction of a scalar field theory.
We demonstrate that these contributions factorize near these zeros for all tree graphs and provide evidence that this extends to loop graphs as well. We explicitly construct polytopal realizations of the relevant graph associahedra and show that the cosmological zeros have natural geometric and physical interpretations.
As a byproduct, we establish an equivalence between the wavefunction coefficients of chain graphs and flat-space Tr$(\phi^3)$ amplitudes, enabling us to leverage the cosmological zeros to uncover the recently discovered hidden zeros of colored amplitudes.
}

\begin{document}

\maketitle

\section{Introduction}

Scattering amplitudes in certain quantum field theories have recently been shown to possess a remarkable property called hidden zeros which are special kinematic configurations of the scattering data for which the amplitudes vanish~\cite{Arkani-Hamed:2023swr}. Furthermore, amplitudes in Tr$(\phi^3)$ theory, the non-linear sigma model, and Yang-Mills theory exhibit a novel factorization property near the kinematical locus of their hidden zeros~\cite{Cachazo:2021wsz, Arkani-Hamed:2023swr, Cao:2024gln, Arkani-Hamed:2024fyd}. While the factorization of an amplitude on its poles is dictated by unitarity, the physical origin of near-zero factorizations and associated splitting behavior remains less understood. Recent progress in this regard has shown that the existence of hidden zeros in these theories is intimately tied to the color-kinematics duality~\cite{Cao:2024gln, Bartsch:2024amu, Li:2024qfp} and enhanced ultraviolet scaling under BCFW shifts~\cite{Rodina:2024yfc}; see also~\cite{Cao:2024qpp, Zhang:2024iun, Li:2024bwq, Zhang:2024efe, GimenezUmbert:2025ech, Huang:2025blb, Backus:2025hpn} for other recent work on hidden zeros, near-zero factorizations and splittings. Earlier examples of amplitude zeros include the Adler zeros of pion amplitudes~\cite{Adler:1964um}, the radiation zero in vector boson pair production~\cite{Dixon:1999di}, and zeros in dual resonant amplitudes studied in the early days of string theory~\cite{DAdda:1971wcy}.

The past few years have also seen considerable progress in our understanding of the analytic structure of observables in cosmological spacetimes from underlying combinatorial and geometric structures. In particular, the Bunch-Davies wavefunction for individual graphs in a theory of conformally coupled scalars can be described combinatorially by graph tubings and geometrically as the canonical form of the cosmological polytope~\cite{Arkani-Hamed:2017fdk} (see~\cite{Benincasa:2022gtd} for a review). More recently, it was shown in~\cite{Arkani-Hamed:2024jbp} that the graph associahedron encodes the combinatorics of graph tubings for single graphs, forming the building blocks for the cosmohedron -- the geometric object capturing sums over graphs contributing to the Tr$(\phi^3)$ wavefunction.

Given the crucial role of geometry (specifically, the ABHY associahedron~\cite{Arkani-Hamed:2017mur}) in revealing hidden zeros of Tr$(\phi^3)$ scattering amplitudes through its ``flattening'' limits, it is natural to explore whether similar features arise in cosmology. In this paper, we investigate hidden zeros of the cosmological Bunch-Davies wavefunction, focusing on flat-space wavefunction coefficients in a theory of massless scalars with a $\phi^3$ interaction, which has been extensively studied in the literature~\cite{Benincasa:2019vqr, Hillman:2019wgh, De:2023xue, Arkani-Hamed:2023bsv, Arkani-Hamed:2023kig, Fan:2024iek, He:2024olr, Benincasa:2024ptf, Baumann:2024mvm, De:2024zic, Du:2024soq}. Unlike the case of flat-space amplitudes, where hidden zeros emerge from full on-shell amplitudes obtained by summing over all Feynman diagrams, we study the structure of hidden zeros in individual diagrams contributing to wavefunction coefficients. By leveraging the geometry of the graph associahedron, we reveal how these zeros manifest in cosmological observables, offering a new perspective on their analytic properties.

The paper is organized as follows. We begin in section~\ref{sec:wavefunc} by establishing notation for the (stripped) wavefunction coefficients corresponding to the various graph topologies (tree-level and one-loop) studied in this work. In section~\ref{sec:cosmologicalzeros} we establish notation for the three different types of cosmological zeros that we will call wavefunction, factorization, and parametric zeros, and work out a couple of examples in detail. We show in section~\ref{sec:graphassoccosmo} that just as the amplitude zeros in Tr$(\phi^3)$ theory can be represented geometrically through various ``flattening'' limits of the associated geometry (the ABHY associahedron) parametric zeros emerge from flattening the geometry that encodes the wavefunction. This geometry, which we call the \textit{cosmological} graph associahedron, is a non-simple polytope that arises from the cosmological limit of the graph associahedron introduced in~\cite{Arkani-Hamed:2024jbp}. We also discuss the geometric and physical origins of the wavefunction and factorization zeros.  In section~\ref{sec:cosmoABHY} we present a surprising connection between the stripped wavefunction coefficients of chain graphs and scattering amplitudes in Tr$(\phi^3)$ theory. This connection remarkably allows us to use the zeros for the wavefunction associated to a single graph to find the zeros of amplitudes in Tr$(\phi^3)$ theory, which involve a sum over many Feynman graphs. We conclude in~\ref{sec:outlook} with a discussion on the scope of our findings and present ideas for future investigations.

\section{(Stripped) wavefunctions}
\label{sec:wavefunc}

Arkani-Hamed, Benincasa and Postnikov showed that the wavefunction coefficients\footnote{Henceforth, we will simply call these \emph{wavefunctions} to avoid clutter.} for a theory of conformally coupled scalars (with polynomial interactions) in an FRW spacetime possess a remarkable combinatorial and geometric description~\cite{Arkani-Hamed:2017fdk} that is independent of Feynman diagrams.
Our focus in this paper will be on the flat-space wavefunctions as their FRW counterparts can be obtained via a simple integral transform.
The combinatorial definition of the contribution $\wc_\G$ to the $n$-site $\ell$-loop\footnote{For loop graphs, $\psi^{(\ell)}_{n}$ is the \textit{loop integrand}.} wavefunction $\wc_n^{(\ell)}$ from a single graph $\G$ proceeds by encircling various subgraphs of $\G$ with tubes to create a graph tubing. A \emph{tube} $\tau$ of $\G$ contains two pieces of information: the set of sites $v$ encircled and the set of edges $e$ crossed.
At tree-level, a tube $\tau$ is unambiguously defined by specifying the sites it encloses, but at loop level one must also specify the edges that are crossed.

If the tube $\tau$ encircles a subgraph with $m$ sites, we refer to it as an $m$-tube.
To compute the wavefunction, we first generate the set of all compatible (non-intersecting) complete tubings $\mathsf{T}$ of $\G$. A \textit{tubing} is a collection of tubes and a \textit{complete tubing} $\mathsf{T}$ is a maximal set of non-overlapping tubes.
We then assign to each complete tubing the inverse of the product of all of the energy factors $S_{\tau}$ associated to the constituent tubings, and sum over all possible compatible complete tubings of $\G$. This leads to a remarkably simple formula for the wavefunction:
\begin{align}\label{eq:wfc}
	\psi_{\G} =
	\sum_\mathsf{T} \prod_{\tau \in  \mathsf{T}}
	\frac{1}{S_\tau}\,,
    \qquad \text{where}
	\qquad
	S_\tau = \sum_{\text{sites } v\in\tau} X_v
	+ \sum_{\text{edges $e$ that cross $\tau$}} Y_e
	\,,
\end{align}
where $X_v$ is the sum of the external energies at site $v$ and $Y_e$ is the energy exchanged through the internal edge $e$ of $\G$.

For a given graph $\G$, a certain subset of tubes is common to all complete tubings $\mathsf{T}$ and hence give factors common to every term in the corresponding expression for the wavefunction.
Therefore, we define the \textit{stripped} wavefunction $\swc_{\G}$, which is associated to a maximal collection of compatible tubes, $\tilde{\mathsf{T}}$, that do not include the total energy $n$-tube $S_{\text{total}}$ (the unique tube encircling all sites and no edges) or the 1-tubes $S_1, \dots ,S_n$ that encircle a single site:
\begin{align}
    \wc_{\G}
    = \frac{1}{S_1 \dots S_n S_{\text{total}}} \times \left(\sum_{\tilde{\mathsf{T}}} \prod_{\tau \in\tilde{\mathsf{T}}} \frac{1}{S_\tau} \right)
    = \frac{1}{S_1 \dots S_n S_{\text{total}}} \times \swc_{\G}\,.
    \label{eq:swfc}
\end{align}
Just like the wavefunction $\wc$, its stripped counterpart $\swc$ also has an independent geometric and combinatorial interpretation:
it was recently shown that $\swc_{\G}$ is the canonical function of the graph associahedron $\A_\G$ whose face structure reflects the combinatorics of the tubings $\tilde{\mathsf{T}}$~\cite{Arkani-Hamed:2024jbp}.
While often non-simple, the cosmological limit of this graph associahedron, $\tilde{\A}_\G$, better reflects the physical properties of the wavefunction (see section~\ref{sec:graphassoccosmo}).
In the remainder of this work, we will focus on the properties of the stripped wavefunction $\swc$ and deduce which of these carry over to the full wavefunction $\wc$.

In the next three subsections we introduce three distinct graph topologies relevant for wavefunctions in a theory with a cubic interaction: the $n$-chain graph and a specific class of $n$-star graphs at tree level, as well as the one-loop $n$-gon graph. We establish notation for the constituent energy factors $S_\tau$ that appear in the various graph topologies with explicit examples.

\subsection{\texorpdfstring{The $n$-chain}{The n-chain}}

For $n$-chains, we use the following convention to label the site and edge energies $X_v$ and $Y_e$:
\begin{center}
\includegraphics[align=c, scale=1.8]{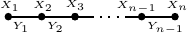}\,.
\end{center}
Then, given a tube $\tau=\{i,i+1,\dots, i+k\}$ of the $n$-chain, the corresponding energy factor $S_\tau$ is given via~\eqref{eq:wfc} by
\begin{align} \label{eq:Schain}
        S_{i\,i{+}1\dots i{+}k}
        = \includegraphics[align=c, scale=1.5]{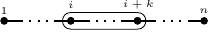}
        &= X_{i} + \cdots + X_{i+k} + Y_{i-1} + Y_{i+k}\,,
\end{align}
where $Y_0 = Y_n = 0$ and the total energy $n$-tube is $S_{\text{total}} = S_{12\dots n} = \sum_{i=1}^n X_i$.
To compute the stripped wavefunction of the $n$-chain, we list its complete compatible tubings $\tilde{\mathsf{T}}$ and use~\eqref{eq:swfc}. For example, the stripped wavefunction for the 3-chain is simply
\begin{align}\label{eq:3chain}
    \swc_{3\ch}
    = \includegraphics[align=c, scale=1.1]{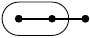}
    ~+~ \includegraphics[align=c, scale=1.1]{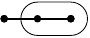}
    = \frac{1}{S_{12}} + \frac{1}{S_{23}}\,,
\end{align}
while the expression for the 4-chain is
\begin{align}
        \swc_{4\ch}
        &= \includegraphics[align=c, scale=.75]{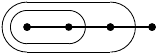}
        		+ \includegraphics[align=c, scale=.75]{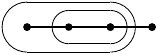}
        		+ \includegraphics[align=c, scale=.75]{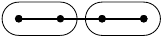}	
		+ \includegraphics[align=c, scale=.75]{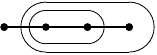}	
		+ \includegraphics[align=c, scale=.75]{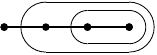}	
        \nn\\
        &= \frac{1}{S_{123}}\left(\frac{1}{S_{12}} + \frac{1}{S_{23}}\right) +\frac{1}{S_{12} S_{34}}
            +\frac{1}{S_{234}}\left(\frac{1}{S_{23}} + \frac{1}{S_{34}}\right) \nn \\
        &= \frac{1}{S_{123}} \swc_{3\ch}(1,2,3) + \frac{1}{S_{12} S_{34}} + \frac{1}{S_{234}} \swc_{3\ch}(2,3,4)\,,
        \label{eq:4chainrecursive}
\end{align}
where the numbers in brackets denote the labels of the ordered sites of the chain subgraphs.
As we have advertised on the last line of~\eqref{eq:4chainrecursive}, it is clear that for general $n$, the stripped wavefunction exhibits a  recursive structure allowing it to be expressed in terms of chain subgraphs as
\begin{align}
\label{eq:n-chain}
    \swc_{n\ch} = \sum_{i=1}^{n{-}1} \frac{1}{S_{1\cdots i}S_{i{+}1\cdots n}}\swc_{i\ch}(1,\cdots, i) \times \swc_{n{-}i\ch}(i{+}1,\cdots, n)\,,
\end{align}
except that the factor $1/S_1$ should be omitted in the $i=1$ term and the factor $1/S_{n}$ should be omitted in the $i=n-1$ term. In applying this formula we note that $\swc_{1\ch} = \swc_{2\ch} =1$.

\subsection{\texorpdfstring{The $n$-star}{The n-star}}

Next, we consider the $n$-star defined and labeled as follows:
\begin{center}
\includegraphics[align=c, scale=1.7]{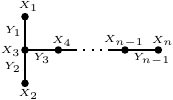}
\,.
\end{center}
This graph can be arranged by connecting an $(n{-}2)$-chain and two 2-chains to a single central site with valence three.
As with chain graphs, we specify a tube $\tau$ by the list of sites encircled: $\tau = \{i_1, \dots, i_k\} \subset \{1, \dots, n\}$ ordered such that $i_j < i_{j+1}$.
The energy factors $S_\tau$ depend on which subset of the sites $\{1,2,3\}$ are included in the tube:
\begin{align} \label{eq:Sstar}
    S^*_{134 \dots k} = X_1 + X_{3} + \cdots + X_{k} + Y_{2} + Y_k\,, &~~~~
    S^*_{234 \dots k} = X_2 + X_{3} + \cdots + X_{k} + Y_{1} + Y_k\,, \nn \\
    S^*_{34 \dots k} = X_{3} + \cdots + X_{k} + Y_{1} + Y_{2} + Y_k\,, &~~~~
    S^*_{1234 \dots k} = X_{1} + \cdots + X_{k} + Y_k \,, \\
    S_{i\,i+1 \dots k} = X_{i} + \cdots& + X_{k} + Y_{i-1} + Y_{k}
    \,, \quad \forall ~ i\geq 4\,, \nn
\end{align}
where we have introduced the notation $S^*$ for energy factors associated with tubings exclusive to the $n$-star, with no analog for $n$-chains. Note that the $S_{i\,i+1 \dots k}$ for $i\geq 4$ are simply the tubes of an $(n{-}3)$-chain~\eqref{eq:Schain} and the total energy tube is that of the $n$-chain, i.e. $S_{\text{total}} = S_{12\dots n} = \sum_{i=1}^n X_i$.
For example, the stripped wavefunction for the 4-star can be expressed in the above notation as
\begin{align}
    \swc_{4\text{-star}}
    &=  \includegraphics[align=c, scale=.85]{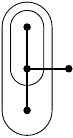}	
    ~+~ \includegraphics[align=c, scale=.85]{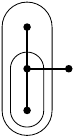}
    ~+~ \includegraphics[align=c, scale=.85]{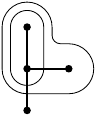}
    ~+~ \includegraphics[align=c, scale=.85]{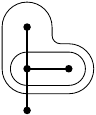}
    ~+~ \includegraphics[align=c, scale=.85]{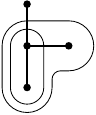}
    ~+~ \includegraphics[align=c, scale=.85]{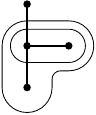}
    \nn\\
    &= \frac{1}{S^*_{123}} \left(\frac{1}{S^*_{13}} {+} \frac{1}{S^*_{23}} \right) + \frac{1}{S^*_{134}} \left(\frac{1}{S^*_{13}} {+} \frac{1}{S^*_{34}} \right) + \frac{1}{S^*_{234}} \left(\frac{1}{S^*_{23}} {+} \frac{1}{S^*_{34}} \right)\,,\nn \\
    &= \frac{1}{S^*_{123}} \swc_{3\ch}(1,3,2) + \frac{1}{S^*_{134}}
    \swc_{3\ch}(1,3,4)
    + \frac{1}{S^*_{234}} \swc_{3\ch} (2,3,4)\,.
\label{eq:111starwfc}
\end{align}
Like with chain topologies, the stripped wavefunction of the general $n$-star can be constructed recursively; this time, in terms of both  chain and star subgraphs
\begin{align}
    \label{eq:nsite_star}
    \tilde{\psi}_{n\text{-star}} &= \frac{1}{S^*_{123} S_{45\dots n}} \tilde{\psi}_{3\text{-chain}}(1,3,2) \times  \tilde{\psi}_{n-3\text{-chain}}(4,\dots, n) +
\sum_{i=1}^2 \frac{1}{S^*_{i3\cdots n}} \tilde{\psi}_{n-1\text{-chain}}(i,3,\dots,n) \, \nonumber \\
&+ \sum_{i=1}^{n-4}\frac{1}{S^*_{123\dots n-i} S_{n-i+1\dots n}} 
  \tilde{\psi}_{n-i\text{-star}}(1,\dots, n-i)
 \times 
  \tilde{\psi}_{i\text{-chain}}(n-i+1,\dots, n)~,
\end{align}
except that all $1/S_n$ factors should be omitted in the  above expression.
The seed of this recursive formula is the expression~\eqref{eq:111starwfc} for the 4-star, which gives a recursive relation for the $n$-star purely in terms of chain subgraphs.

\subsection{\texorpdfstring{The $n$-gon}{The n-gon}}

For graphs corresponding to loop integrands, we only need to consider the 1-loop $n$-gon since any other graph can be obtained by gluing trees to the $n$-gon.
We arrange the $n$ sites of the $n$-gon on a circle in a clockwise manner:
\begin{center}
    \includegraphics[align=c, scale=1.5]{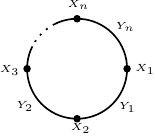}
    \,.
\end{center}
The indices labeling the sites of the $n$-gon are to be interpreted cyclically. The energy tubes $S_{\tau}$ contributing to $\swc_{n\text{-gon}}$ are specified by a list of cyclically consecutive sites $\tau=\{i,i+1,\dots, j\}$
\begin{align}\label{eq:ngonSij}
    S_{i\,i+1 \dots j} &= \includegraphics[align=c, scale=1.1]{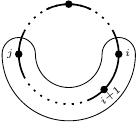}
    = X_i + \cdots + X_j + Y_{i-1} + Y_{j}
    \,,
\end{align}
for all tubes other than the total energy tube $S_{\text{total}}$.
One should think of the tubes in~\eqref{eq:ngonSij} as starting at $i$ and ending at $j$; in particular, tubes that start at $i$ and end at $j=i{-}1$, i.e. $\tau =\{i, i+1, \dots, i-1\}$ also contribute to the stripped wavefunction.
We label the total energy tube by $\tau=\{1,2,\dots, n, 1\} = \{i, i+1, \dots, n, 1, \dots, i\}$ and draw it as:
\begin{align} \label{eq:ngonStot}
    S_\text{total}
    = S_{12 \cdots n1}
    &= \includegraphics[align=c, scale=1.1]{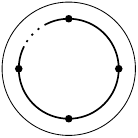}
    = \sum_{i=1}^n X_i
    \,.
\end{align}
Notice that both tubes $\tau=\{1, 2, \dots, n\}$ and $\tau=\{1, 2, \dots, n, 1\}$ include all sites.
However, the two are distinct -- the former crosses the edge labeled by $Y_n$ while the latter does not cross any edge.
This is because at loop-level specifying only the set of sites enclosed is is not enough to uniquely fix a tube.
To keep the notation light for $n$-gons, we fix this ambiguity by including the site labeled $1$ twice in the total energy tube.
One should think of this tube as starting and ending at the same site.

Using the notation we have introduced, the 3-gon can for example be expressed as
\begin{align}
    \swc_{3\text{-gon}}
    &= \includegraphics[align=c, scale=.6]{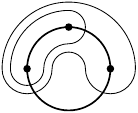}
	+ \includegraphics[align=c, scale=.6]{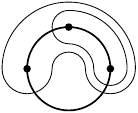}
	+ \includegraphics[align=c, scale=.6]{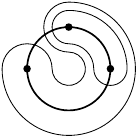}
	+ \includegraphics[align=c, scale=.6]{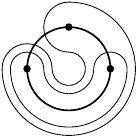}
	+ \includegraphics[align=c, scale=.6]{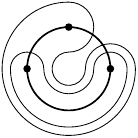}
	+ \includegraphics[align=c, scale=.6]{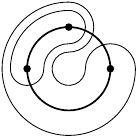}
    \nn\\
    &= \frac{1}{S_{123}}\left(\frac{1}{S_{12}}+\frac{1}{S_{23}}\right)+\frac{1}{S_{231}}\left(\frac{1}{S_{23}}+\frac{1}{S_{31}}\right)+\frac{1}{S_{312}}\left(\frac{1}{S_{31}}+\frac{1}{S_{12}}\right)\,, \nn \\
    &= \frac{1}{S_{123}} \swc_{3\ch}(1,2,3) + \frac{1}{S_{231}} \swc_{3\ch}(2,3,1) + \frac{1}{S_{312}} \swc_{3\ch}(3,1,2)\,.
   \label{eq:3gonwfc}
\end{align}
The recursive pattern demonstrated in the last line above persists for general $n$-gons and the corresponding stripped wavefunction can also be written for any $n$ solely in terms of the $n$-chain via
\begin{align}
    \swc_{n\text{-gon}} = \sum_\sigma \frac{1}{S_\sigma} \swc_{n\ch} (\sigma[1,\cdots,n])\,,
\label{eq:ngonswc}
\end{align}
where $\sigma$ is a cyclic permutation of the site labels $\{1,\cdots,n\}$.

\paragraph{Arbitrary graphs.} For an arbitrary tree- or loop-level graph one can systematically derive formulas analogous to~\eqref{eq:n-chain}, \eqref{eq:nsite_star}, and~\eqref{eq:ngonswc}, so that ultimately any stripped wavefunction can be constructed recursively by ``gluing'' chains together.

\section{Cosmological zeros}
\label{sec:cosmologicalzeros}

In this section we introduce three categories of linear conditions on the energy factors for which (stripped) wavefunctions vanish that we call wavefunction zeros (section~\ref{sec:wav0}), factorization zeros (section~\ref{sec:fac0}) and parametric zeros (section~\ref{sec:par0}). In sections~\ref{sec:example5star} and~\ref{sec:example3gon} we exhibit these types of zeros in two detailed examples.

\subsection{Wavefunction zeros}
\label{sec:wav0}

We begin with a class of zeros that is possessed by the chain graphs only. One can easily verify that for $n>2$, the wavefunction of the $n$-chain obtained from~\eqref{eq:wfc} vanishes when the following $n{-}2$ conditions are satisfied:
\begin{align}
\label{eq:nchain_wfzero}
    S_{12\cdots n{-}1} - S_{1\cdots i}+S_{i{+}1\cdots n}=0 \text{ for all }  2\le i\le n{-}2 \quad \text{and} \quad  S_{12\cdots n{-}1}+ S_{23\cdots n}=0\,.
\end{align}
We call these \emph{wavefunction zeros} because these are zeros of both the full wavefunction $\wc_{n\ch}$ and its stripped counterpart $\swc_{n\ch}$.

In fact the $n$-chain wavefunction satisfies a universal factorization property near its wavefunction zero.\footnote{This is analogous to the \textit{near-zero} factorization associated to amplitudes zeros discovered in \cite{Arkani-Hamed:2023swr} where all but one of its zero conditions is relaxed.}
Consider a kinematic configuration $\mathcal{S}_i$ in which all of the conditions~\eqref{eq:nchain_wfzero} are satisfied except for a single $i \in \{2,\ldots,n-2\}$
\begin{align}
    S_{12\cdots n{-}1}- S_{1 \cdots i}+S_{i+1\cdots n}\ne 0\,.
    \label{eq:relaxwfzero}
\end{align}
Then one can easily check that the $n$-chain wavefunction $\wc_{n\ch}$ factors into a product of chains obtained by ``cutting'' the $n$-chain between sites $i$ and $i{+}1$:
\begin{align}
    \wc_{n\ch} \rvert_{\mathcal{S}_{i}}= \frac{S_{12\cdots n{-}1}- S_{1\cdots i}+S_{i{+}1\cdots n}}{S_{12\cdots n}S_{12\cdots n{-}1}} \times \psi_{i\ch}(1,{\cdots},{i}) \times \psi_{n{-}i\ch}({i{+}1},{\cdots},n)\,.
    \label{eq:splittingtheorem1}
\end{align}
This factorization property makes manifest the existence of a zero at the locus~\eqref{eq:nchain_wfzero}.
When interpreting the equation~\eqref{eq:splittingtheorem1} and several similar ones that follow, it is crucial to keep in mind that the energy factors involving the ``cut'' $i$-th and $(i{+}1)$-th sites in the subgraphs on the right-hand side are identified with those that appear in the parent $n$-site graph on the left; in other words, despite being exterior sites of the subgraph, $i$ and $i{+}1$ still carry the same interior energy factors as before. Let us illustrate this point explicitly by way of an example for $n=4$ and $i=2$ where we look at the locus
\begin{align}
\mathcal{S}_{2} = \{S_{123} + S_{234} = 0 \}
\,.
\end{align}
A short calculation reveals that
\begin{align}
\psi_{4\ch}\rvert_{\mathcal{S}_{2}} &= \left.\includegraphics[align=c,scale=0.8]{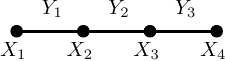}
\right|_{\mathcal{S}_{2}}\nn\\
&= \frac{S_{123}-S_{12}+S_{34}}{S_{1234} S_{123}} \times \includegraphics[align=c,scale=0.8]{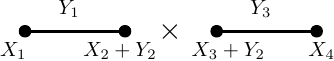}\nn\\
    &=  \frac{S_{123}-S_{12}+S_{34}}{S_{1234} S_{123}} \times \psi_{2\ch}(1,{2}) \times \psi_{2\ch}({3},4)\,,
\end{align}
where, as indicated by the pictures on the second line, we have
\begin{align}
\psi_{2\ch}(1,{2}) &=\frac{1}{S_1S_2S_{12}} = \frac{1}{(X_1 + Y_1)(X_2+Y_1+Y_2)(X_1+X_2+Y_2)}\,, \\
\psi_{2\ch}({3},4) &= \frac{1}{S_3S_4S_{34}} = \frac{1}{(X_3+Y_2+Y_3)(X_4+Y_3)(X_3+X_4+Y_2)}\,.
\end{align}

A similar factorization property exists on the locus $\mathcal{S}$ on which all of the conditions in~\eqref{eq:nchain_wfzero} except that $S_{12\cdots n{-}1}+ S_{23\cdots n} \ne 0$. In this case,
\begin{align}
    \wc_{n\ch}\rvert_{\mathcal{S}} &= \frac{S_{12\cdots n{-}1}+S_{23\cdots n}}{S_{12\cdots n}S_{12\cdots n{-}1}}\times \psi_{1\ch}({1}) \times \psi_{n{-}1\ch}({2},3,\cdots,n)\\
    &= \frac{S_{12\cdots n{-}1}+S_{23\cdots n}}{S_{12\cdots n}S_{23\cdots n{-}1}} \times \psi_{n{-}1\ch}(1,2,\cdots,{n{-}1})\times \psi_{1\ch}({n})\,,
    \label{eq:splittingtheorem2}
\end{align}
where the symmetry of the chain under reversal of the labeling is reflected in the existence of two equivalent formulas; each manifests the wavefunction zero on the locus~\eqref{eq:nchain_wfzero}. Again, we note that the wavefunctions on the right-hand side of~\eqref{eq:splittingtheorem2} must be properly interpreted in terms of shifted energy variables. For example, at $n=4$ we have
\begin{align}
    \wc_{4\ch}\rvert_{\mathcal{S}} &=
    \frac{S_{123}+S_{234}}{S_{1234} S_{123}} \times \includegraphics[align=c,scale=0.8]{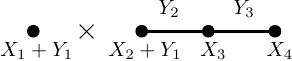}\nn \\
    &= \frac{S_{123}+S_{234}}{S_{1234} S_{123}} \times \psi_{1\ch}({1}) \times \psi_{3\ch}({2},3,4)~\,,
    \label{eq:5sitewavefunczerofactorization2}
\end{align}
where in this case
\begin{align}
\psi_{1\ch}({1}) &= \frac{1}{S_1} = \frac{1}{X_1 + Y_1}\,,\\
\qquad
\psi_{3\ch}({2},3,4) &= \frac{\frac{1}{S_{23}} + \frac{1}{S_{34}}}{S_2S_3S_4S_{234}} = \frac{\frac{1}{X_2+X_3+Y_1+Y_3} + \frac{1}{X_3+X_4+Y_2}}{(X_2 + Y_1 + Y_2)(X_3+Y_2+Y_3)(X_4+Y_3)(X_2+X_3+X_4+Y_1)}\,.
\end{align}

\subsection{Factorization zeros}
\label{sec:fac0}

It is well-known that wavefunctions obey factorization on the residues of their physical poles at $S_{\tau \in \mathsf{T}} = 0$~\cite{Benincasa:2019vqr, Arkani-Hamed:2018kmz, Baumann:2021fxj}, with the residue on the total energy pole at $S_\text{total} = 0$ famously being the corresponding flat-space S-matrix element.
Note that a subset of the energy factors, namely the ones $\{ S_1, \dots, S_n, S_\text{total} \}$ that belong to tubes in $\mathsf{T}$ but not in $\tilde{\mathsf{T}}$ and arise from the prefactor in~\eqref{eq:swfc}, are not poles of stripped $n$-site wavefunctions.
We will refer to these energy factors as the \emph{parameters} of a stripped $n$-site wavefunction.
In this section, we demonstrate that stripped wavefunctions also exhibit factorization when any one of the parameters $S_j$ corresponding to an interior site\footnote{An interior site is one that is adjacent to at least two other sites. Specifically, in an $n$-chain the sites labeled $i \in \{2,\dots,n-1\}$ are interior, in an $n$-star the sites labeled $i \in \{3,\dots,n-1\}$ are interior, while in an $n$-gon all sites are interior. The remaining sites are called exterior.} is set to zero. These factorizations imply that the stripped wavefunction associated to any graph must exhibit zeros associated to its various subgraphs; we call these \emph{factorization zeros}.\footnote{Consequently, the factorization associated with factorization zeros requires relaxing a larger set of conditions (i.e., those corresponding to the zeros of the involved subgraphs) closely resembling the \textit{splitting} phenomenon observed in amplitudes zeros \cite{Cachazo:2021wsz, Cao:2024gln}. Nevertheless, the resulting expressions for the factorized stripped wavefunctions take a form similar to that obtained by relaxing a single condition in a parametric zero (section \ref{sec:par0}). In this sense, the factorization associated with factorization zeros can be viewed as interpolating between splitting and near-zero factorization in the terminology introduced for amplitudes zeros.}

\paragraph{$n$-chain.} For any $j \in \{2,\ldots,n-1\}$ it follows from~\eqref{eq:n-chain} and the definition~\eqref{eq:Schain} after a little bit of work that
\begin{align}
\label{eq:n-chainsplit}
    \tilde\psi_{n\ch}\rvert_{S_j = 0}=\frac{S_{12\cdots n}}{S_{1\cdots j}S_{j\cdots n}} \times \tilde\psi_{j\ch}(1,\cdots, {j}) \times \tilde\psi_{n{-}j{+}1\ch}({j},\cdots, n)\,.
\end{align}
As always, following the discussion below~\eqref{eq:splittingtheorem1}, we emphasize that the energy factors at site $j$ on the right-hand side are the same as in the parent graph on the left-hand side.
We note that~\eqref{eq:n-chainsplit} explains why the 3-chain has no factorization zeros: on the support of the only possible factorization condition $S_2=0$, the 3-chain factorizes into a product of two 2-chains, which do not have any zeros of their own since $\swc_{2\ch}=1$.

\paragraph{$n$-star.} For the $n$-star there are two distinct types of factorization limits depending on whether we factor at the trivalent site or (for $n>4$) at one of the other $n-4$ interior sites. In the former case, it follows from~\eqref{eq:nsite_star} that
\begin{align}
\label{eq:starfact1}
\swc_{n\text{-star}}\rvert_{S^*_3 = 0} &=\frac{S_{12\cdots n}}{S_{13}^* S_{23}^* S^*_{3\cdots n}} \times \swc_{2\ch} (1,{3}) \times \swc_{2\ch} (2,{3}) \times  \swc_{n{-}2\ch}({3},4,\cdots,n) \,,
\end{align}
while in the latter case we have for $j \in \{4,\ldots,n{-}1\}$ that
\begin{align}
\label{eq:starfact2}
\swc_{n\text{-star}}\rvert_{S_j = 0} &=\frac{S_{1234\cdots n}}{S_{j\cdots n}S^*_{1234\cdots j}} \times \swc_{j\text{-star}} (1,\cdots,{j}) \times \swc_{n{-}j{+}1\ch}({j},\cdots,n)\,.
\end{align}
We note that the 4-star has no factorization zeros since the only factorization~\eqref{eq:starfact1} leads to a product of three 2-chains which are each just $\swc_{2\ch}=1$.

\paragraph{$n$-gons.} Using the recursion~\eqref{eq:ngonswc} one can see for any $i \in \{1,\ldots,n\}$ the $n$-gon ``factors'' as
\begin{align}
    \swc_{n\text{-gon}}\rvert_{S_i=0}= \swc_{n{+}1\ch} ({i},i{+}1,\cdots,n,1,2,\cdots,{i})\,.
    \label{eq:ngonfactSicond2}
\end{align}

\paragraph{Arbitrary graphs.} Any zero of a stripped wavefunction appearing on the right-hand side of a factorization can be considered a ``factorization zero'' of the parent stripped wavefunction.
However, in order to enumerate all of the distinct zeros arising from an arbitrary tree- or loop-level graph, it is sufficient to look at its chain subgraphs, after using the fact that any graph can be broken into chains by taking some sequence of $S_i \to 0$ limits at various interior sites $i$.
Moreover, for any tree graph, tabulating the wavefunction zeros of all chain subgraphs is sufficient to give the full set of distinct factorization zeros due to a cancellation of total energy factors in intermediate steps.
These phenomena are exhibited by the examples worked out in detail in sections~\ref{sec:example5star} and~\ref{sec:example3gon} below.

\subsection{Parametric zeros}
\label{sec:par0}

It is evident that the right-hand side of the factorization~\eqref{eq:n-chainsplit} of the $n$-chain vanishes when the total energy factor $S_{12 \dots n}$ is zero.
The same is true for the $n$-star as seen from~\eqref{eq:starfact1} and~\eqref{eq:starfact2}. Finally, by taking the $S_j \to 0$ limit on both sides of the factorization~\eqref{eq:ngonfactSicond2}, it is clear that the $n$-gon inherits the zeros of the chain appearing on the right-hand side of~\eqref{eq:ngonfactSicond2}.  We refer to zeros of this type as \emph{parametric zeros} since they involve simply setting certain parameters of the stripped wavefunction to zero.
In summary, parametric zeros for the three relevant graph topologies are given by:
\begin{align}
    \label{eq:parChain}
    \text{$n$-chain: } & \quad
    S_i = 0 \text{ and } S_{\text{total}} = 0
    \text{ for any } i \in \{2,\dots, n{-}1\}
    \,,
    \\
    \label{eq:parStar}
   \text{$n$-star: }  & \quad
    S_3^* = 0 \text{ and } S_{\text{total}} = 0
    \,,
    \quad \text{ or } \quad S_{i}=0 \text{ and } S_{\text{total}} = 0
   \text{ for any }  i \in \{4,\dots, n{-}1\}
    \,,
    \\
    \label{eq:parGon}
    \text{$n$-gon: } & \quad
    S_i = 0 \text{ and } S_j = 0 \text{ and } S_{\text{total}} = 0
    \text{ for any }  i, j \in \{1,\dots, n\}
    \text{ and }
    i \neq j
    \,.
\end{align}

\subsection{Example: the 5-star}
\label{sec:example5star}

The stripped wavefunction $\swc_{5\text{-star}}$ of the 5-star has no wavefunction zeros (since only chains have wavefunction zeros) and, according to~\eqref{eq:parStar}, two parametric zeros: at $S_3^* = S_{12345} = 0$ and at $S_4 = S_{12345} = 0$. Let us now classify its factorization zeros, which arise from two possible factorizations: at $S_3^* = 0$ or at $S_4 = 0$.

On the support of $S_3^* = 0$ (which we denote graphically by drawing a dashed circle around site 3), \eqref{eq:starfact1} tells us that the stripped wavefunction factors into a certain prefactor times the product of its constituent chain subgraphs, two of which are trivial due to $\swc_{2\ch} = 1$:
    \begin{align}
    \swc_{5\text{-star}} \rvert_{S_3^* = 0}
    =\includegraphics[align=c,scale=0.7]{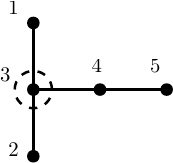} &= \frac{S_{12345}}{S_{13}^* S_{23}^* S_{345}^*} \times 1 \times 1 \times \swc_{3\ch}({3},4,5)\,.
\end{align}
This vanishes if we further impose the wavefunction zero condition~\eqref{eq:nchain_wfzero} for the 3-chain subgraph $\swc_{3\ch}({3},4,5)$ given by the relation
\begin{equation}
    S_{{3}4}^* + S_{45} = 0\,.
\end{equation}
Therefore, $S_3^*=S_{{3}4}^* + S_{45}=0$ constitutes a factorization zero of $\swc_{5\text{-star}}$.

On the other hand, on the support of $S_4=0$, \eqref{eq:starfact2} tells us that
    \begin{align}
    \swc_{5\text{-star}} \rvert_{S_4 = 0}
    = \includegraphics[align=c,scale=0.7]{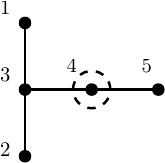} &= \frac{S_{12345}}{S_{1234}^* S_{45}} \times \swc_{4\text{-star}}(1,2,3,{4}) \times 1\,.
    \label{eq:5sitestarfactS4}
    \end{align}
Now $\swc_{4\text{-star}}$ does not possess any wavefunction zeros (since only chains have those), and it does not possess any factorization zeros (see the comment under~\eqref{eq:starfact2}); the only possible path to zero is via the parametric zero of the 4-star $\swc_{4\text{-star}}(1,2,3,{4})$ at $S^*_3=S_{1234}^*=0$.  However, on the support of $S^*_{3}=0$, the stripped wavefunction of the 4-star subgraph further factors into
\begin{align}
    \swc_{5\text{-star}}\rvert_{S_3^*=S_{4}=0}
    = \frac{S_{12345}}{S_{1234}^* S_{45}} \times \includegraphics[align=c,scale=0.72]{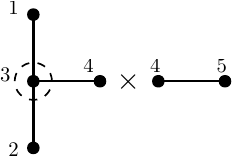} = \frac{S_{12345}}{S_{13}^* S_{23}^* S_{34}^* S_{45}} \times \includegraphics[align=c,scale=0.72]{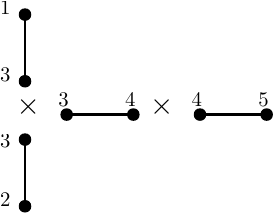}\,.
    \label{eq:5sitestarfactS3S4}
    \end{align}
Note in the second equality the factor $S_{1234}^*$ has canceled out and therefore the second condition of the parametric zero is rendered useless. With only 2-chains remaining in the product, we find no factorization zeros of $\swc_{5\text{-star}}$ on the locus $S_4 = 0$. Note that had we started with an $(n{>}5)$-star graph, we would have reached an $(n{-}3)$-chain, and its wavefunction zeros would have led to factorization zeros of $\swc_{(n{>}5)\text{-star}}$ in this step.

\subsection{Example: the 3-gon}
\label{sec:example3gon}

The stripped wavefunction $\swc_{3\text{-gon}}$ of the 3-gon has no wavefunction zeros (since only chains have wavefunction zeros) and three parametric zeros, at the loci given by~\eqref{eq:parGon}.  Let us now classify its factorization zeros. According to~\eqref{eq:ngonfactSicond2} it ``factors'' at $S_1 = 0$ into the 4-chain
\begin{align}
    \swc_{\text{3-gon}} \rvert_{S_1=0} =
    \includegraphics[align=c,width=.12\textwidth]{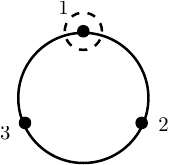}
    = \includegraphics[align=c,width=.16\textwidth]{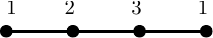}
    = \swc_{4\ch}({1},2,3,{1})\,.
    \label{eq:3gonS1fact}
\end{align}
The 4-chain appearing here has parametric zeros, but these are the same as the parametric zeros of the 3-gon itself that we already listed.
In addition, the 4-chain vanishes at the wavefunction zero given by~\eqref{eq:nchain_wfzero}.
Since we could have factored at sites 2 or 3, altogether we obtain three distinct factorization zeros of the 3-gon, at the loci
\begin{align}
    S_1 &= S_{123}+S_{231} = S_{12}-S_{31}+S_{231} = 0\,,\\
    S_2 &= S_{231}+S_{312} = S_{23}-S_{12}+S_{312} = 0\,, \\
    S_3 &= S_{312}+S_{123} = S_{31}-S_{23}+S_{123} = 0\,.
\end{align}
These arguments straightforwardly generalize to the $n$-gon, which inherits the wavefunction zeros of the $(n{+}1)$-chain on the locus $S_i=0$ for any $i$.

\section{The geometric origins of cosmological zeros}
\label{sec:graphassoccosmo}

In section~\ref{sec:AG}, we construct in kinematic space the graph associahedron associated to a stripped wavefunction and discuss a particular cosmological limit that degenerates the graph associahedron into a (generically non-simple) polytope relevant to our analysis of cosmological zeros.
Through explicit examples in sections~\ref{sec:pzerosnondegen} and~\ref{sec:pzerosdegen}, we show that a graph associahedron can be expressed in terms of its constituent Minkowski summands; moreover, we demonstrate that tuning the parameters in a particular manner causes the cosmological graph associahedron to flatten (i.e., the polytope drops in dimension), thereby providing a geometric origin for the parametric zeros of section~\ref{sec:par0}.
In section~\ref{sec:geo-origins}, we discuss the geometric origin of the wavefunction and factorization zeros of sections~\ref{sec:wav0} and~\ref{sec:fac0} through the lens of the vanishing locus of the adjoint polynomial.

\subsection{Graph associahedra and the cosmological limit}
\label{sec:AG}

It has been recently shown~\cite{Arkani-Hamed:2024jbp} that the stripped wavefunction $\swc_{\G}$ associated to a single Feynman graph $\G$ is the canonical form of a polytope called a graph associahedron $\A_\G$ (slightly different than the one used in the mathematical literature),
whose facet structure reflects the combinatorics of the tubings $\tilde{\mathsf{T}}$ of $\G$.
These graph associahedra are facets of larger polytopes called cosmohedra~\cite{Arkani-Hamed:2024jbp}, which encode the full flat-space wavefunctions (summed over all contributing Feynman diagrams) of a colored version of the uncolored theory studied in this paper at the level of individual diagrams.
In this section we provide an algorithm for constructing graph associahedra based on simplex truncation introduced by Carr and Devadoss in~\cite{carr2005coxetercomplexesgraphassociahedra, devadoss2006realizationgraphassociahedra}; a complementary construction was recently given in~\cite{Glew:2025otn, Glew:2025ugf}.

To construct the graph associahedron associated to a single graph $\G$ relevant for the cosmological context, we modify the Carr-Devadoss algorithm so that 2-tubes replace 1-tubes as the fundamental building blocks.
This is natural because the energy factors of 1-tubes only appear as overall factors in the wavefunction associated to $\G$.
Also, our construction will heavily rely on the fact that the energy factors satisfy the linear relations
\begin{align}
\label{eq:linRel}
    S_{\tau_1 } + S_{\tau_2}
    = S_{\tau_1 \cup \tau_2} + S_{\tau_1 \cap \tau_2}\,,
\end{align}
when $\tau_1 \cap \tau_2$ is non-empty.
Note that beyond tree-level one must be careful in interpreting this formula since the order of the indices that define a tube matters.
For example, already at one-loop $\tau_1$ can intersect $\tau_2$ twice -- at both ends of $\tau_1$ -- and the operation $\cap$ can only be thought of as acting on the tube not the indices labeling a tube.

\begin{figure}
    \centering
    \includegraphics[width=0.8\linewidth]{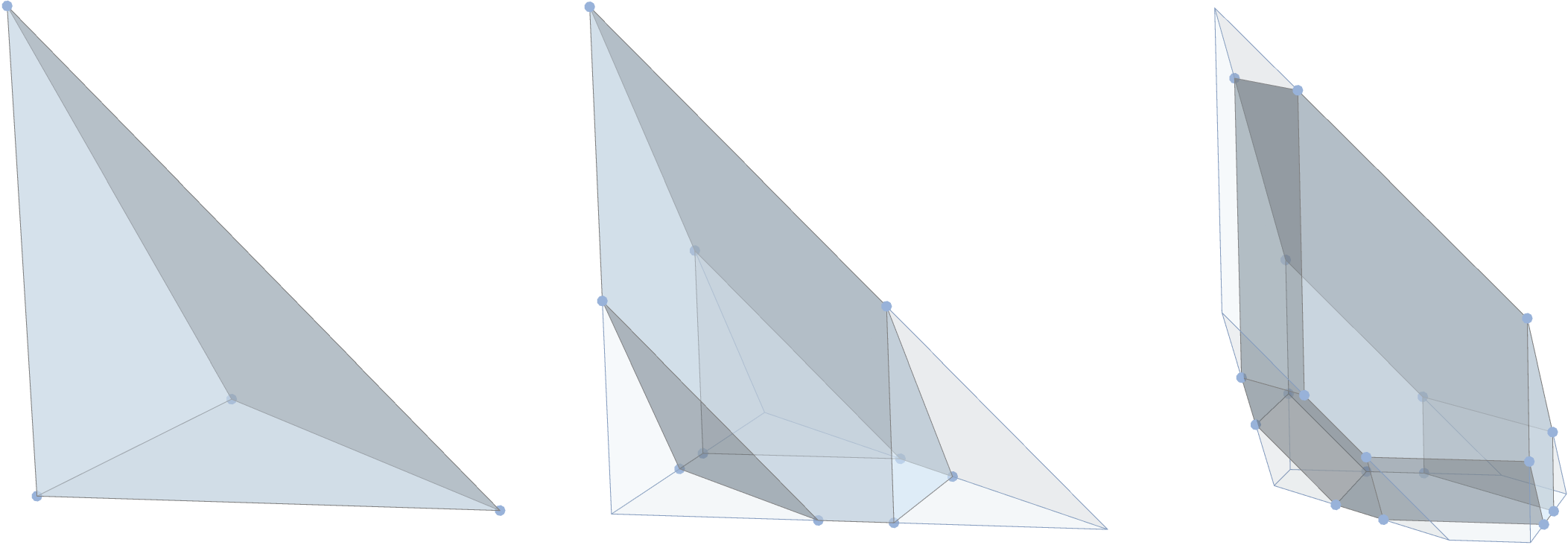}
    \caption{Truncating a 3-simplex to carve out the graph associahedron $\A_{5\text{-star}}$.
    Left: we begin with the simplex cut out by the 2-tube hyperplanes $\{S^*_{13}, S^*_{23}, S^*_{34}, S_{45}\}=0$.
    Middle: the vertices of the simplex are truncated by the 4-tube hyperplanes $\{S^*_{1234}, S^*_{1345}, S^*_{2345}\}=0$.
    Right: the edges of the simplex are truncated by the 3-tube hyperplanes $\{S^*_{123}, S^*_{134}, S^*_{234}, S^*_{345}\}=0$.
    }
    \label{fig:simplexTruncation}
\end{figure}

Let $\mathcal{T}_{k}$ denote the set of all $k$-tubes of some graph $\G$ and let $m=|\mathcal{T}_{2}|= n{+}\ell{-}1$ be the total number of 2-tubes.
The graph associahedron $\A_\G$ can be realized as a simple $(m{-}1)$-dimensional polytope in an $m$-dimensional ambient space $\mathbb{R}^m$ with coordinates $S_{\tau \in \mathcal{T}_2}$ labeled by the 2-tubings.
To construct $\A_\G$ we start with the $(m{-}1)$-simplex (see for example figure~\ref{fig:simplexTruncation} left)
defined by intersecting $\{S_\tau \ge 0: \tau \in \mathcal{T}_2\} \subset \mathbb{R}^m$ with the hyperplane
\begin{align}
	\sum_{\tau \in \mathcal{T}_2} S_{\tau}
        = S_\text{total}
        + \sum_{\text{interior sites}\, v} \Big(c_v(\tau_\text{total})-1\Big)\,  S_v
	\,,
\end{align}
where $\tau_\text{total}$ is the total energy tube\footnote{Recall that $\tau_\text{total} = \{1,\cdots, n\}$ for  chains and stars while $\tau_\text{total} = \{1,\cdots, n,1\}$ for $n$-gons.}, the  second sum  includes only the interior sites of $\G$,
and $c_v(\tau)$ is the number of 2-tubes in $\tau$ that include the interior site $v$.
Then, for each $k \in \{3,\ldots,n+\ell-1\}$, each $(m{-}k)$-dimensional facet of the simplex is truncated by the hyperplanes $S_{\tau \in \mathcal{T}_k}=0$ where $S_\tau$ is defined by
\begin{align} \label{eq:deformedLinRel}
	\sum_{\tau^\prime \in \mathcal{T}_2 \,:\, \tau^\prime  \subset \tau} S_{\tau^\prime}
	= S_{\tau}
        + \sum_{\text{interior sites\,} v} \Big(c_v(\tau)-1\Big) S_v
        - \delta_{\tau}
	\,,
\end{align}
where the $\delta_\tau$ are positive parameters whose role will be explained shortly.
See figure~\ref{fig:simplexTruncation} for a visualization of this process.
Also note that while the 2-tubes are naturally distinguished by the Carr-Devadoss construction, one could choose any cardinality-$m$ subset of the energy variables $S_{\tau}$ (for $\tau \in \tilde{\mathsf{T}})$ as coordinates for the ambient space.

\begin{figure}
\begin{center}
\includegraphics[width=.5\textwidth]{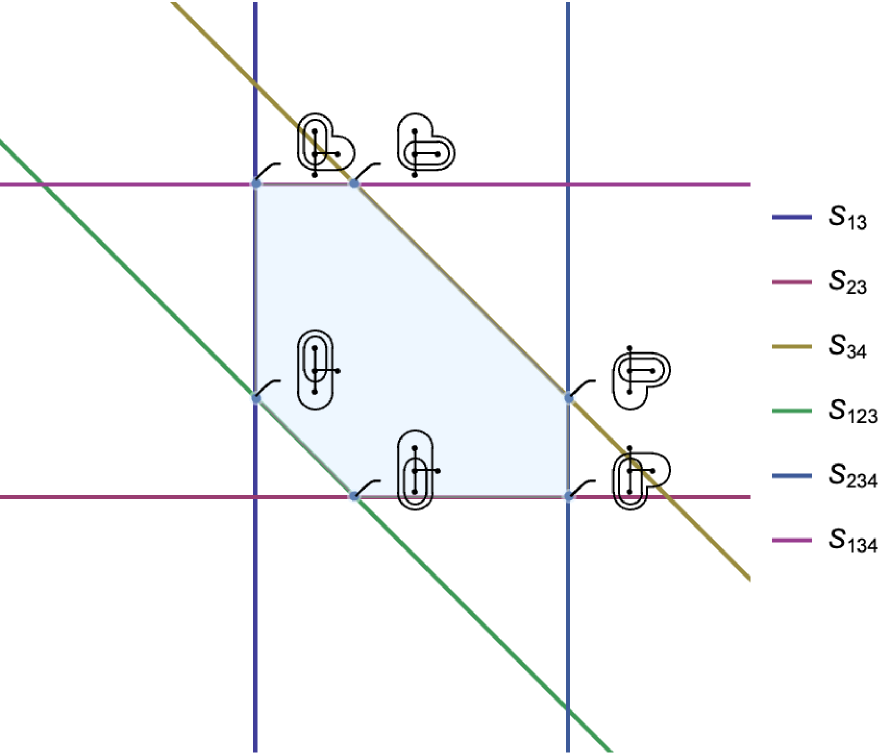}
\qquad
\includegraphics[width=.4\textwidth]{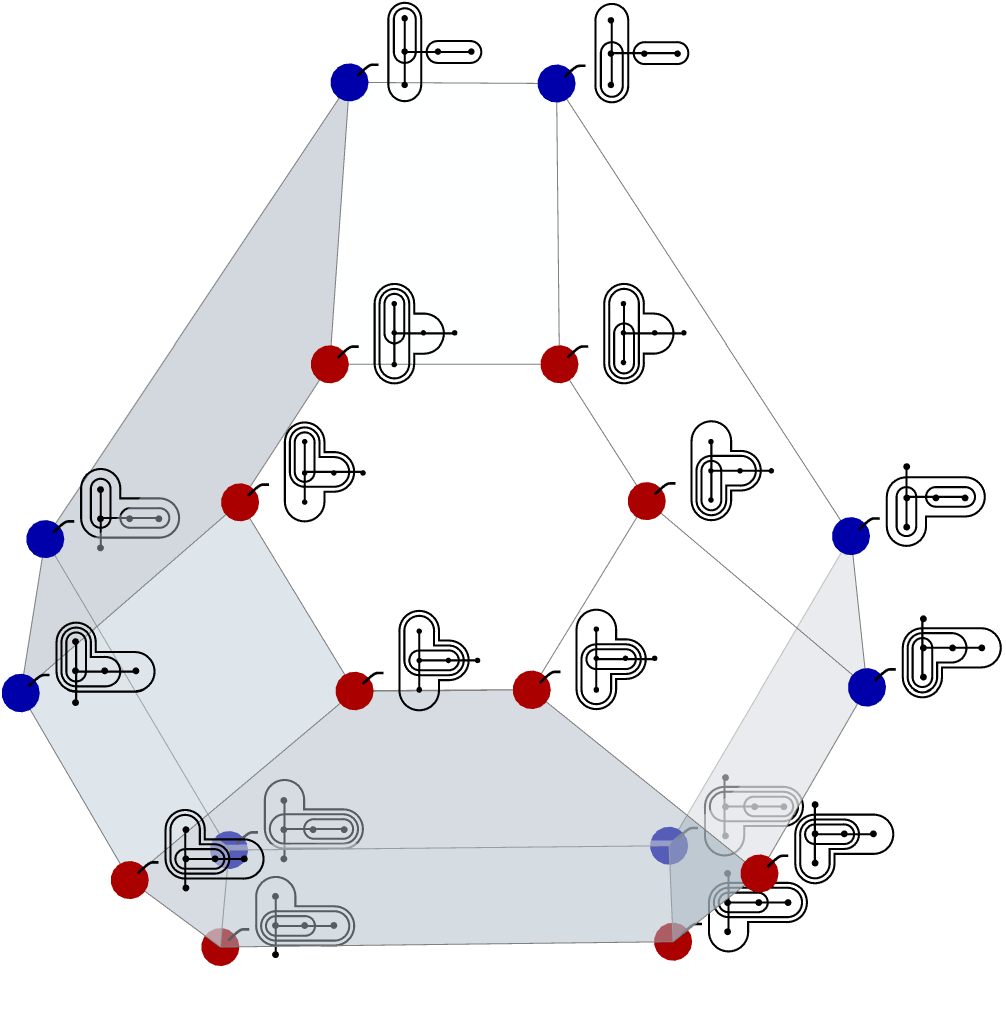}
\end{center}
\caption{%
The graph associahedra $\A_{4\text{-star}}$ (left) and $\A_{5\text{-star}}$ (right), with vertices labeled by complete tubings.}
\label{fig:deformedPolytopes}
\end{figure}

Now, in order to avoid cutting the simplex too deeply and to ensure that the resulting polytope remains simple, we require that the $\delta_\tau$'s introduced in~\eqref{eq:deformedLinRel} satisfy the bounds
\begin{align} \label{eq:deltacond1}
	\delta_{\tau} > 0 \text{ for all } \tau
	\quad \text{and} \quad
	\delta_{\tau_1} + \delta_{\tau_2}
	> \delta_{\tau_1 \cup \tau_2} + \delta_{\tau_1 \cap \tau_2} \text{ when } \tau_1\cap \tau_2 \ne \emptyset\,,
\end{align}
as well as
\begin{align} \label{eq:deltacond2}
    0 <	\delta_{\tau\in \mathcal{T}_3} < S_v < S_\text{total}
    \text{ for all internal sites } v
    \,.
\end{align}
The boundary of the region defined by $\{ S_{\tau \in \mathcal{T}_k} \geq 0: 2 \le k \le n - 1\}$ then provides a realization of the graph associahedron $\A_\G$ as the boundary of an $(m{-}1=n{+}\ell{-}2)$-dimensional polytope.
Each complete compatible tubing $\tilde{\mathsf{T}}$ of the stripped wavefunction $\swc_\G$ corresponds to a
vertex of $\A_\G$ where $\cap_{\tau \in \tilde{\mathsf{T}}} \{S_\tau = 0\}$.
Figure~\ref{fig:deformedPolytopes} depicts the graph associahedra for the 4- and 5-star graphs.

\begin{figure}
\begin{center}
\includegraphics[width=.9
\textwidth]{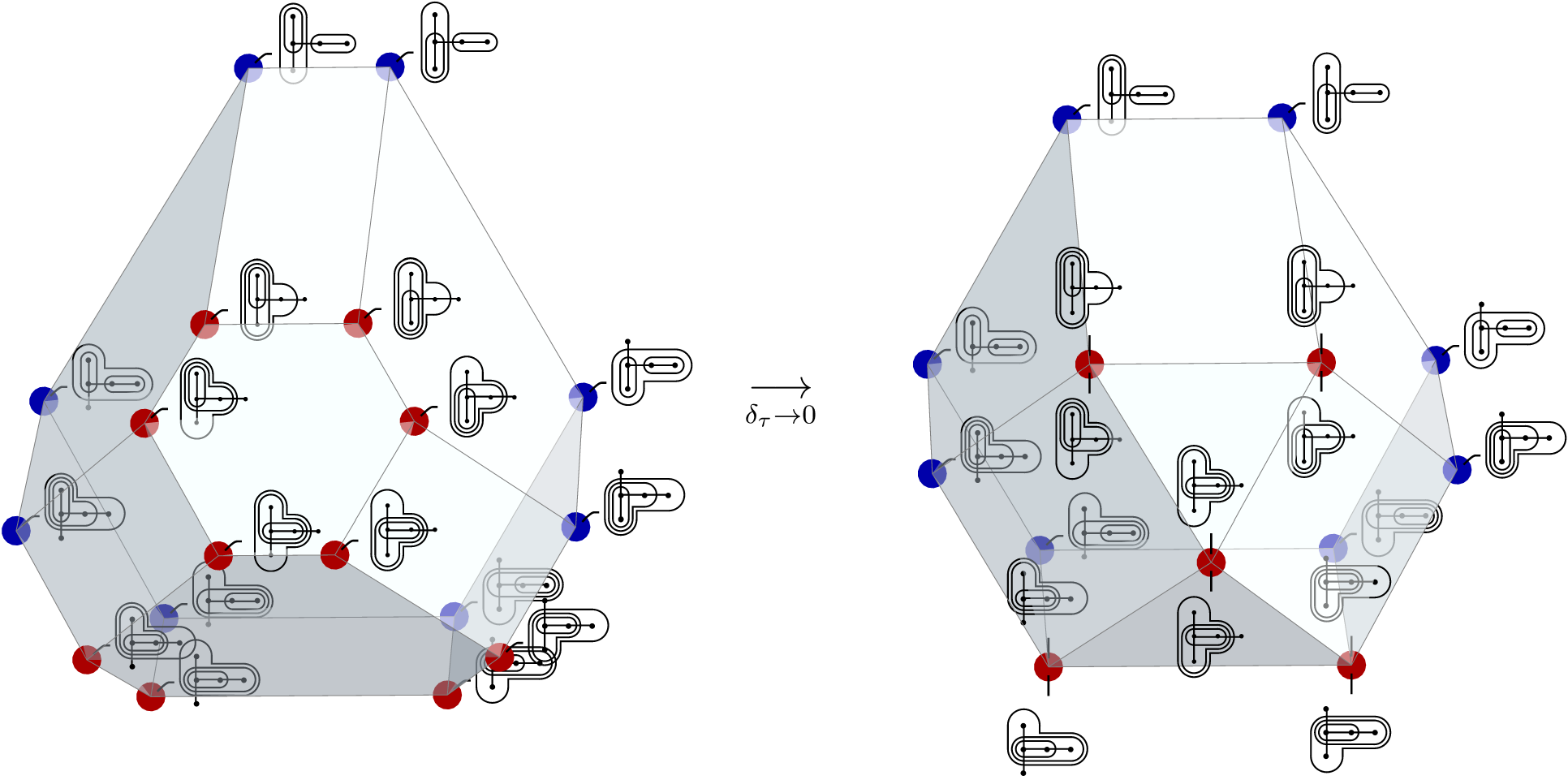}
\end{center}
\caption{%
Left: the graph associahedron $\A_{5\text{-star}}$ ($\delta_\tau \neq 0$); the $f$-vector of this simple polytope is (18,27,11). Right: the cosmological graph associahedron $\tilde{\A}_{5\text{-star}}$ ($\delta_\tau = 0$); the $f$-vector of this non-simple polytope is (13,22,11). Note that as expected, the number of facets remains the same, while the number of edges and vertices decreases. The vertices that remain simple in the $\delta_\tau \to 0$ limit are colored {\color{MidnightBlue}blue}, while all others are colored {\color{BrickRed}red}.
}
\label{fig:112star}
\end{figure}

By construction, the non-zero $\delta_\tau$ ensure that the graph associahedron ${\A}_\G$ is always a simple polytope.
While $\A_\G$ does describe the combinatorics of compatible tubings of $\G$,
the energy factors defined by~\eqref{eq:deformedLinRel} do not in general satisfy the relations~\eqref{eq:linRel} required for the cosmological application.
This requires taking the limit $\delta_\tau \to 0$, which results in a non-simple polytope when $n+\ell>4$.
We will call the resulting
polytope $\tilde{\A}_\G := \lim_{\delta_\tau\to0} \A_\G$ the \emph{cosmological graph associahedron} associated to $\G$.
Figure~\ref{fig:112star} shows the degeneration from the simple $\delta$-deformed graph associahedron to the
non-simple cosmological graph associahedron for the 5-star graph.

Even though we are primarily interested in the cosmological limit $\delta_\tau \to 0$, the polytope $\A_\G$ is an important intermediate step that encodes useful combinatorial information.
Moreover, the canonical function of the cosmological  graph associahedron $\tilde{\A}_{\G}$, defined by taking the $\delta_\tau\to0$ limit of the canonical function of the simple polytope $\A_\G$, is precisely the stripped wavefunction $\swc_\G$.
Note that this limit is automatically realized when the energy factors $S_\tau$ are parameterized directly in terms of the energy variables $X_v$ and $Y_e$ of section~\ref{sec:wavefunc}.

\subsection{Parametric zeros: examples of simple polytopes}
\label{sec:pzerosnondegen}

In this section, we illustrate how the parametric zeros of section~\ref{sec:par0} can be seen as flattening limits of the cosmological graph associahedron for three examples -- the 4-chain, the 4-star and the 3-gon.
For these examples $n+\ell = 4$ and, consequently, both $\A_\G$ and $\tilde{\A}_\G$ are simple polytopes that are topologically equivalent.
Nevertheless they are geometrically distinct, having different Minkowski summands and flattening behaviors.
Aside from selected comments, we set the $\delta_\tau$ to zero from the beginning and explore the properties of $\tilde{\A}_\G$ in the following.

\begin{figure}
    \centering
    \includegraphics[scale=0.75]{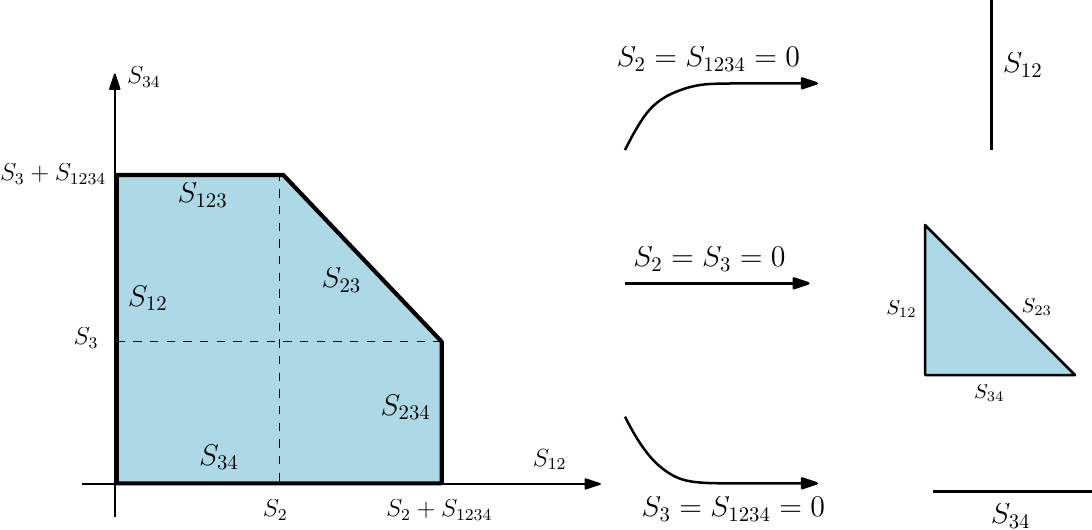}
    \caption{The cosmological graph associahedron $\tilde{\A}_{4\ch}$ and its Minkowski summands, showing that the polytope collapses in the upper and lower limits, corresponding to parametric zeros of $\swc_{4\ch}$, but not in the center limit.}
    \label{fig:4sitegraphassockin}
\end{figure}

\paragraph{The 4-chain.}
Following section~\ref{sec:AG}, we carve out the cosmological graph associahedron $\tilde{\A}_{4\ch}$ in kinematic space with coordinates $(S_{12}, S_{34})$ and parameters $(S_2, S_3, S_{1234})$ via the inequalities
\begin{equation}
\begin{gathered}
    S_{12} \geq 0\,,\qquad S_{34} \geq 0\,,\qquad S_{23} = S_2 + S_3 + S_{1234} - S_{12} -S_{34} \geq 0\,, \\
    S_{123} = S_3 + S_{1234} - S_{34} \geq 0 \,,\qquad S_{234} = S_2 + S_{1234} - S_{12} \geq 0\,,
\end{gathered}
\end{equation}
which describes a pentagon.
This polytope and its Minkowski summands are presented in figure~\ref{fig:4sitegraphassockin}.
From~\eqref{eq:parChain}, we expect the associated stripped wavefunction $\swc_{4\ch}$ to have two parametric zeros, at
\begin{align}
\label{eq:parchain4}
S_{2} = S_{1234} = 0 \qquad \text{ and at } \qquad  S_{3} = S_{1234} = 0\,.
\end{align}
As shown in the top and bottom right of figure~\ref{fig:4sitegraphassockin}, these limits evidently collapse the pentagon to a vertical or horizontal line segment, respectively.
In both cases the stripped wavefunction $\swc_{4\ch}$ vanishes and the polytope $\tilde{\A}_{4\ch}$ collapses in dimension to one of its 1-dimensional Minkowski summands.
On the other hand, the limit $S_2 = S_3 = 0$ collapses the pentagon to its 2-dimensional Minkowski summand -- the triangle (center right of figure~\ref{fig:4sitegraphassockin}).
Since the polytope does not drop in dimension, this limit does not correspond to a zero of the stripped wavefunction.

\begin{figure}
    \centering
    \includegraphics[scale=0.75]{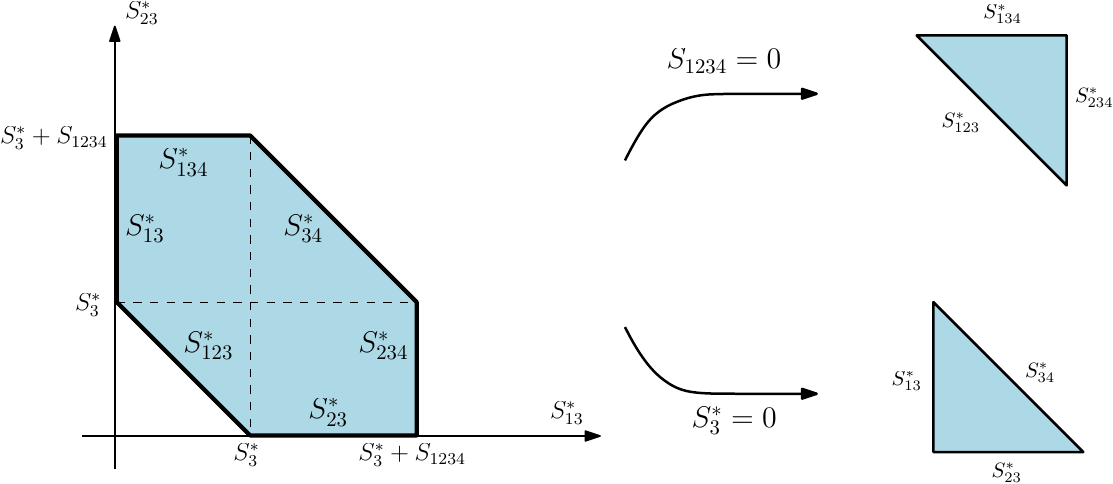}
    \caption{The cosmological graph associahedron $\tilde{\A}_{4\text{-star}}$ and its Minkowski summands, showing that the polytope does not collapse in either of the limits shown.}
    \label{fig:4sitestargraphassockin}
\end{figure}

\paragraph{The 4-star.}
Again following the prescription of section~\ref{sec:AG} and choosing coordinates $(S^*_{13}, S^*_{23})$ with $(S^*_3, S_{1234})$ as parameters,
the cosmological graph associahedron $\tilde{\A}_{4\text{-star}}$ is the hexagon in figure~\ref{fig:4sitestargraphassockin} carved out by the inequalities
\begin{equation}
\begin{gathered}
    S^*_{13} \geq 0\,, \qquad S^*_{23} \geq 0\,, \qquad S^*_{34} = S_{1234} {+} 2S_3^* {-} S^*_{13} {-} S^*_{23}
    \geq 0\,, \\
    S^*_{134} = S^*_{13} {+} S^*_{34} {-} S_3^* \geq 0\,,\qquad S^*_{123} =
    S^*_{13} {+} S^*_{23} {-} S_3^* \geq 0\,, \qquad
    S^*_{234} = S^*_{23} {+} S^*_{34} {-} S_3^* \geq 0\,.
\end{gathered}
\end{equation}
We expect the sole zero of $\swc_{4\text{-star}}$ to be the parametric zero at (see~\eqref{eq:parStar})
\begin{align}
S_3^* = S_{1234} =0\,.
\label{eq:soleparamzero4star}
\end{align}
Note that the 4-star has only two parameters in contrast to three for the 4-chain.
Setting either of them to zero collapses the hexagon to one of its two Minkowski summands -- the triangles shown on the right of figure~\ref{fig:4sitestargraphassockin}.
Since the polytope does not drop in dimension, these limits do not correspond to zeros of $\swc_{4\text{-star}}$.
The parametric zero~\eqref{eq:soleparamzero4star} corresponds to collapsing the hexagon down to a point -- a 2-dimensional drop.

The above observations are simple consequences of the linear relations~\eqref{eq:linRel} and the cosmological limit.
The $\delta$-deformed graph associahedron $\A_{4\text{-star}}$ behaves more like $\tilde{\A}_{4\ch}$ since the nonzero $\delta_\tau$ behave as extra parameters.
Indeed, one can find flattening limits of $\A_{4\text{-star}}$ that only drop the dimension of the polytope by one instead of two.

\begin{figure}
    \centering
    \includegraphics[scale=0.75]{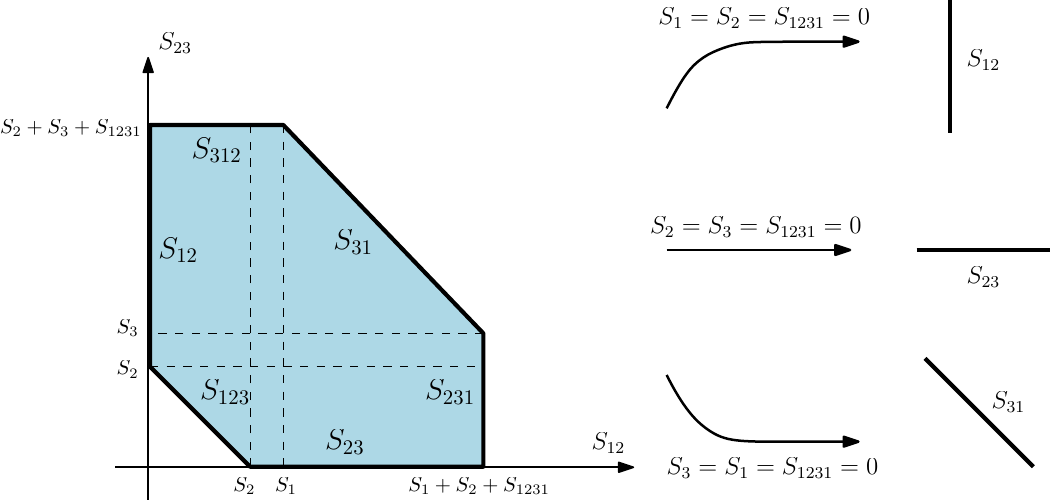}
    \caption{The graph associahedron $\tilde{\A}_{3\text{-gon}}$ and its Minkowski summands. Note that here we only show the lower-dimensional Minkowski summands, not the full Minkowski decomposition of the hexagon. In particular we do not show the summand that corresponds to the limit $S_1=S_2=S_3=0$, since that leaves a 2-dimensional simplex.}
    \label{fig:3gongraphassockin}
\end{figure}

\paragraph{The 3-gon.}
Our final example of a simple cosmological  graph associahedron is $\tilde{\A}_{3\text{-gon}}$.
Choosing coordinates $(S_{12}, S_{23})$ with parameters ($S_1, S_2, S_3, S_{1231}$), the 3-gon inequalities are
\begin{equation}
\begin{gathered}
    S_{12} \geq 0\,,\qquad S_{23} \geq0\,, \qquad
    S_{31} = S_{1231} + S_1 + S_2 + S_3 - S_{12}-S_{23} \geq 0\,, \\
    S_{123} = S_{12}+S_{23}-S_2 \geq 0\,, \quad S_{231} = S_{23} + S_{31} - S_3
    \geq 0\,, \quad S_{312} = S_{31}+S_{12}-S_1 \geq 0
    \,.
\end{gathered}
\end{equation}
The inequalities again carve out a hexagon, as shown in figure~\ref{fig:3gongraphassockin}.
From~\eqref{eq:parGon} we expect that $\swc_{3\text{-gon}}$ as three parametric zeros, at
\begin{align}
    S_1=S_2=S_{1231}=0\,,
    &&
    S_2=S_3=S_{1231}=0\,,
    && \text{and }
    S_3=S_1=S_{1231}=0\,,
\end{align}
corresponding to flattening limits of the hexagon to 1-dimensional Minkowski summands as shown in figure~\ref{fig:3gongraphassockin} (right).
Interestingly, note that although both $\tilde{\A}_{4\text{-star}}$ and $\tilde{\A}_{3\text{-gon}}$ are hexagons, the existence of two extra parameters in the latter implies that its associated Minkowski summands are 1-dimensional line segments.

\subsection{Parametric zeros: examples of non-simple polytopes}
\label{sec:pzerosdegen}

Starting in 3-dimensions the distinction between the graph associahedron $\A_\G$ and its cosmological limit $\tilde{\A}_{\G}$ becomes more apparent: the former is simple while the latter is not.

\begin{figure}
\centering
\includegraphics[scale=0.55]{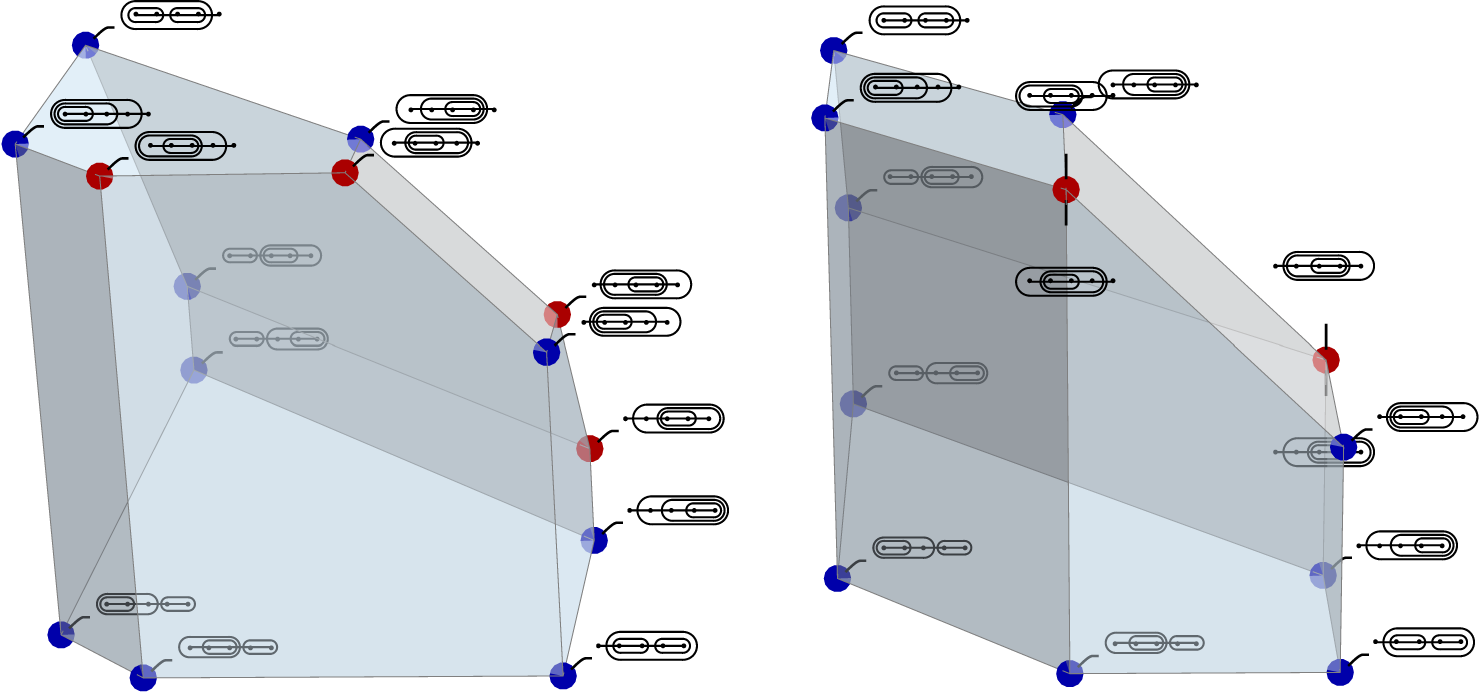}
\caption{%
    The polytopes $\A_{5\ch}$ (simple with $f$-vector (14,21,9)) and $\tilde{\A}_{5\ch}$ (non-simple with $f$-vector (12,19,9)) for the 5-chain graph.
    The {\color{MidnightBlue}blue} vertices remain simple in the cosmological limit $\delta_\tau\to 0$ while the {\color{BrickRed}red} vertices become non-simple.
}
\label{fig:5sitechainassoc}
\end{figure}

\paragraph{The 5-chain.}
Choosing $(S_{12}, S_{23}, S_{45})$ as coordinates with parameters $(S_2, S_3, S_4, S_{12345})$, the simple polytope $\A_{5\ch}$ is defined by the inequalities
\begin{equation}
\begin{gathered}
    S_{12} \geq 0\,, \qquad S_{23} \geq 0\,,\qquad S_{45} \geq 0\,, \quad S_{34} {=} S_{12345} {+} S_2 {+} S_3 {+} S_4 {-} S_{12} {-} S_{23} {-} S_{45}  \geq 0\,,\\
    S_{123} =  \delta_{123} {+} S_{12} {+} S_{23} {-} S_{2} \geq 0\,,\quad S_{234} = \delta_{234} {+} S_{23} {+} S_{34} {-} S_{3}
    \geq 0\,, \quad S_{345} = \delta_{345} {+} S_{34} {+} S_{45} {-} S_{4}
    \geq 0\,, \\
    S_{1234} = \delta_{1234} {+} S_{12} {+} S_{23} {+} S_{34} {-} S_2 {-} S_3
    \geq 0\,,\quad S_{2345} = \delta_{2345} {+} S_{12} {+} S_{23} {+} S_{34} {+} S_{45} {-} S_{2} {-} S_{3} {-} S_{4} \geq 0 \,.
\end{gathered}
\end{equation}
The graph associahedron $\A_{5\ch}$  and its cosmological limit $\tilde{\A}_{5\ch}$ are depicted in figure~\ref{fig:5sitechainassoc}.

\begin{figure}[ht]
\centering
\includegraphics[align=c, scale=.5]{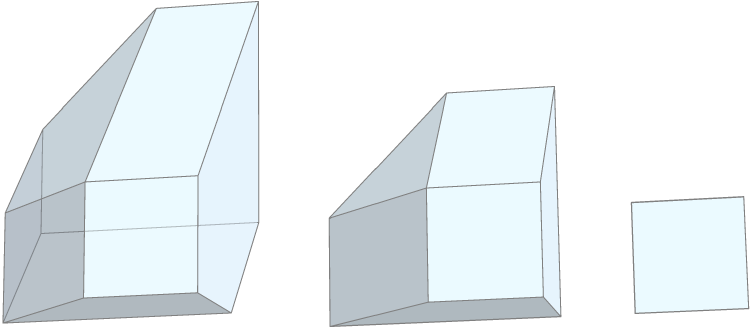}
\qquad\qquad
\includegraphics[align=c, scale=.5]{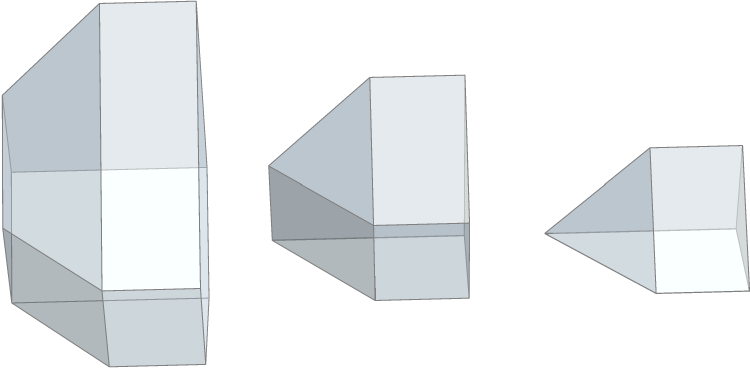}
\caption{%
Left: the sequential limit $\lim_{S_{12345} \to 0} \lim_{S_2 \to 0} \tilde{\A}_{5\ch}$.
The polytope drops in dimension, signaling the parametric zero of $\swc_{5\ch}$.
Right: the sequential limit $\lim_{S_{3} \to 0} \lim_{S_2 \to 0} \tilde{\A}_{5\ch}$.
The polytope does not drop in dimension indicating that this limit does not lead to a parametric zero.
}
\label{fig:5chianParemetricZeros}
\end{figure}

As expected from~\eqref{eq:parChain}, the sequential limit $\lim_{S_{12345}\to0} \lim_{S_{i} \to 0} \tilde{\A}_{5\ch}$ flattens the polytope for $i=2,3,4$.
An example of this is shown in the left of figure~\ref{fig:5chianParemetricZeros}.
On the other hand, the polytope cannot be flattened by keeping the total energy factor non-zero.
For example, the sequential limit $\lim_{S_{3} \to 0} \lim_{S_2 \to 0} \tilde{\A}_{5\ch}$ is still a 3-dimensional polytope (see the right of figure~\ref{fig:5chianParemetricZeros}) and consequently does not correspond to a zero of $\swc_{5\ch}$.

\paragraph{The 5-star.}
In the coordinates $(S^*_{13}, S^*_{23}, S^*_{34})$ with parameters $(S^*_3, S_4, S_{12345})$, the inequalities that define the graph associahedron $\A_{5\text{-star}}$ are
\begin{align}
    S^*_{13} &\geq 0
    \,,
    \hspace{7em}
    S^*_{23}\geq 0
    \,,
    &
    S^*_{34} &\geq 0
    \,,
    \nn\\
    S_{45}&= S_{12345}{+}2 S_{3}^*{+}S_{4} {-}S^*_{13}{-}S^*_{23}{-}S^*_{34} \geq 0
    \,,
    &
    S^*_{123}&= \delta_{123}{+}S^*_{13}{+}S^*_{23}{-}S_{3}^*\geq 0
    \,,
    \nn\\
    S^*_{345}&=\delta_{345}{-}S^*_{13}{-}S^*_{23}{+}S_{12345}{+}2 S_{3}^*\geq 0
    \,,
    &
    S^*_{134}&=\delta_{134}{+}S^*_{13}{+}S^*_{34}{-}S_{3}^*\geq 0
    \,,
    \\
    S^*_{234}&=\delta_{234}{+}S^*_{23}{+}S^*_{34}{-}S_{3}^*\geq 0
    \,,
    &
    S^*_{1234}&=\delta_{1234}{+}S^*_{13}{+}S^*_{23}{+}S^*_{34}{-}2 S_{3}^*\geq 0
    \,,
    \nn\\
    S^*_{1345}&=\delta_{1345}{-}S^*_{23}{+}S_{12345}{+}S_{3}^*\geq 0
    \,,
    &
    S^*_{2345}&=\delta_{2345}{-}S^*_{13}{+}S_{12345}{+}S_{3}^*\geq 0
    \nn\,.
\end{align}
The graph associahedron $\A_{5\text{-star}}$ and its cosmological limit $\tilde{\A}_{5\text{-star}}$ are shown in figure~\ref{fig:112star}.

\begin{figure}[ht]
\centering
\includegraphics[align=c, scale=.5]{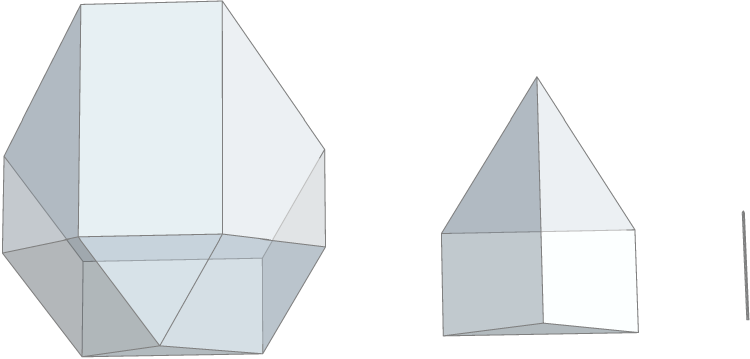}
\qquad\qquad
\includegraphics[align=c, scale=.5]{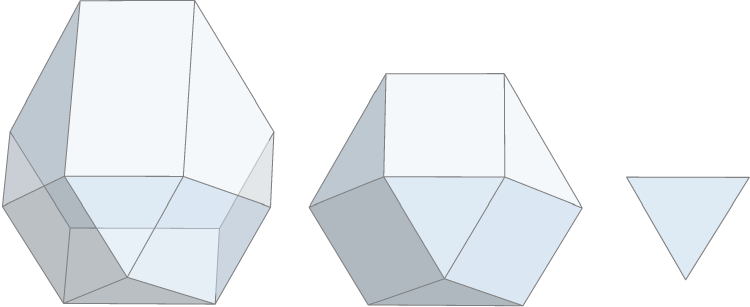}
\caption{%
The sequential limits $\lim_{S_{12345} \to 0} \lim_{S_3^* \to 0} \tilde{\A}_{5\text{-star}}$ (left, showing a 2-dimensional drop) and  $\lim_{S_{12345} \to 0} \lim_{S_4 \to 0} \tilde{\A}_{5\text{-star}}$ (right, showing a 1-dimensional drop). }
\label{fig:5starParemetricZeros}
\end{figure}

The 5-star exhibits two distinct flavors of parametric zeros; both are depicted in figure~\ref{fig:5starParemetricZeros}.
The parametric zero defined by the sequential limit $\lim_{S_{12345} \to 0} \lim_{S_3^* \to 0} \tilde{\A}_{5\text{-star}}$ behaves like that of the 4-star where the dimension of the polytope drops by two: from a 3-dimensional polytope to a 1-dimensional line.
The only other parametric zero, corresponding to the sequential limit $\lim_{S_{12345} \to 0} \lim_{S_4 \to 0} \tilde{\A}_{5\text{-star}}$, behaves like that of the 5-chain since the polytope dimension only drops by one.

\begin{figure}
\centering
\includegraphics[width=\textwidth]{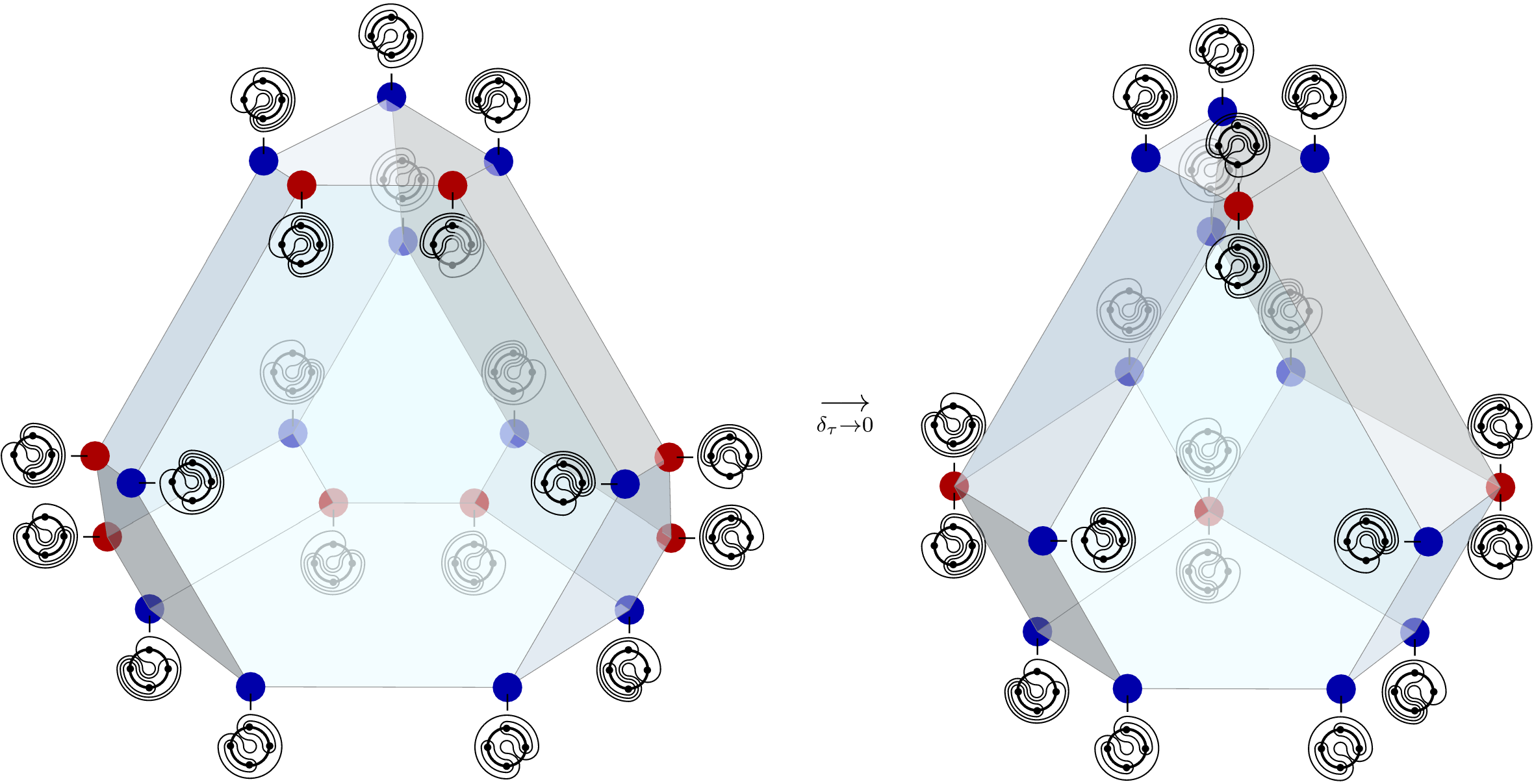}
\caption{%
    The polytopes $\A_{4\text{-gon}}$ (simple with $f$-vector $(20, 30, 12)$) and $\tilde{\A}_{4\text{-gon}}$ (non-simple with $f$-vector $(16, 26, 12)$) for the 4-gon graph.
    The {\color{MidnightBlue}blue} vertices remain simple in the cosmological limit $\delta_\tau\to 0$ while the {\color{BrickRed}red} vertices become non-simple.
}
\label{fig:4gonassoc}
\end{figure}

\begin{figure}
\centering
\includegraphics[width=\textwidth]{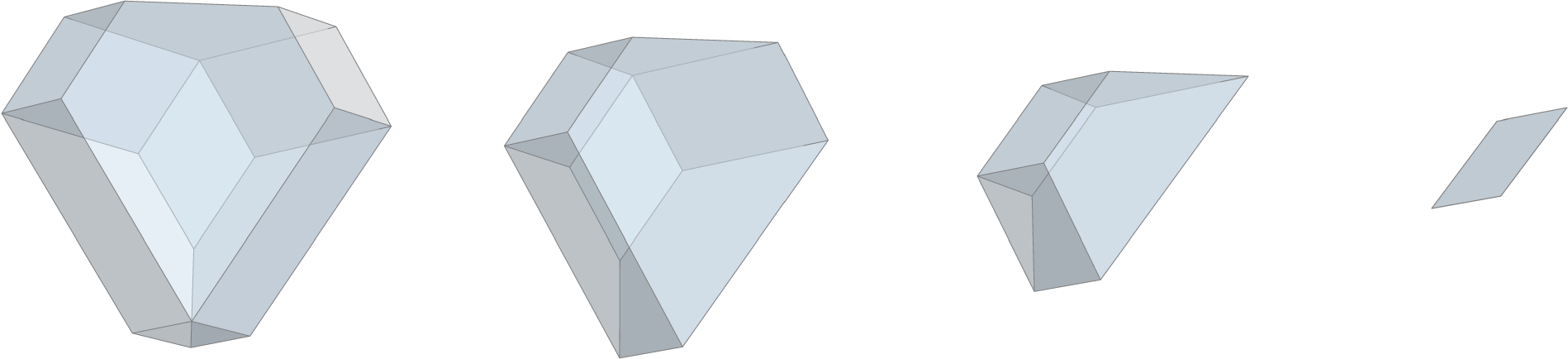}
\caption{%
    The sequential limit $\lim_{S_{12341}\to0} \lim_{S_{2}\to0} \lim_{S_{1}\to0} \tilde{\A}_{4\text{-gon}}$, showing a 1-dimensional drop.
}
\label{fig:4gonFlattening}
\end{figure}

\paragraph{The 4-gon.}
In the coordinates $(S_{12}, S_{23}, S_{34})$ with parameters $(S_1, S_2, S_3, S_4, S_{12341})$, the inequalities that define $\A_{4\text{-gon}}$ are
\begin{align}
    S_{12} &\geq 0\,,
    &
    S_{23} &\geq0
    \,,\nn\\
    S_{34} &\geq0
    \,,
    &
    S_{41} &=S_{12341}{+}S_1{+}S_2{+}S_3{+}S_4{-}S_{12}{-}S_{23}{-}S_{34}\geq0
    \,,\nn\\
    S_{123} &= \delta_{123}{+}S_{12}{-}S_2{+}S_{23} \geq0\,,
    & S_{341} &= \delta_{341}{+}S_{12341}{+}S_2{+}S_1{+}S_3
        {-}S_{12}{-}S_{23}
    \,,\\
    S_{234} &= \delta_{234}{+}S_{23}{-}S_3{+}S_{34} \geq0
    \geq0\,,
    & S_{412} &= \delta _{412}
        {+}S_{12341}{+}S_2{+}S_3{+}S_4
        {-}S_{23}{-}S_{34}
    \geq0
    \,,\nn\\
    S_{1234} &= \delta_{1234}
        {-}S_2{-}S_3
        {+}S_{12}{+}S_{23}{+}S_{34}
    \geq0
    \,, &
    S_{2341} &= \delta _{2341}
        {+}S_{12341}{+}S_1{+}S_2
        {-}S_{12}
    \geq0
    \,,\nn\\
    S_{3412} &= \delta_{3412}
        {+}S_{12341}{+}S_2{+}S_3
        {-}S_{23}
    \geq0
    \,,
    &
    S_{4123} &= \delta_{4123}
        {+}S_{12341}{+}S_3{+}S_4
        {-}S_{34}
    \geq0
    \,.\nn
\end{align}
Both the graph associahedron $\A_{4\text{-gon}}$ and its cosmological limit $\tilde{\A}_{4\text{-gon}}$ are depicted in figure~\ref{fig:4gonassoc}.
The parametric zeros of $\swc_{4\text{-gon}}$ correspond to flattening limits where the dimension of the polytope drops by one.
To illustrate this we demonstrate the flattening corresponding to the parametric zero $S_{1}=S_{2}=S_{12341}=0$ in figure~\ref{fig:4gonFlattening}.

\subsection{Wavefunction and factorization zeros from the adjoint polynomial}
\label{sec:geo-origins}

Unlike the parametric zeros discussed above, the wavefunction zeros of $n$-chains discussed in section~\ref{sec:wav0} (and consequently, the factorization zeros of arbitrary graphs) are not realized as flattening limits of the associated cosmological graph associahedra. Instead, we will now illustrate that they can be understood from a geometric perspective by studying the adjoint polynomial of the associated polytope; in particular, we will see that while all of the parametric zeros are accessible from within the positive region (by dialing various positive parameters to zero), the factorization and wavefunction zeros lie outside of it.  We utilize the 4-chain as an illustrative example throughout this section.

\begin{figure}
    \centering
    \includegraphics[align=c,scale=0.45]{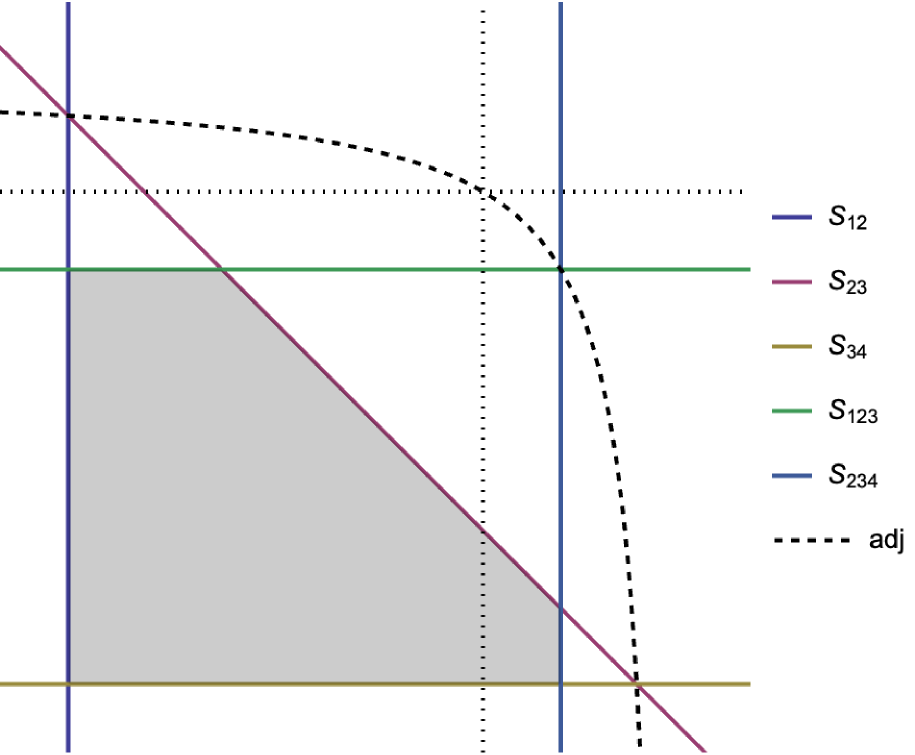}
    \quad
    \includegraphics[align=c,scale=0.45]{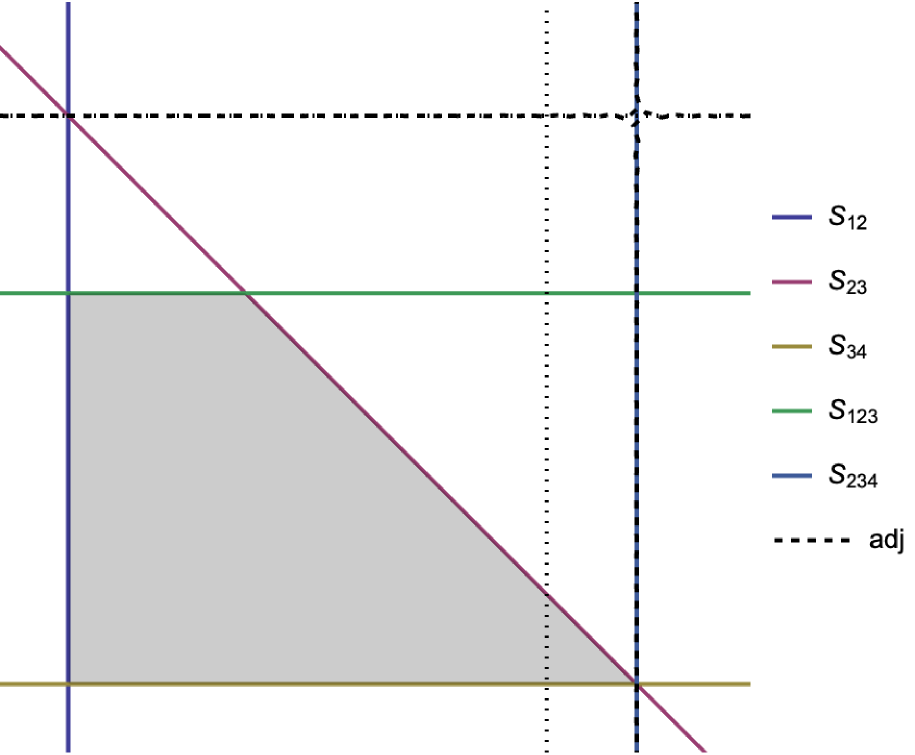}
    \caption{%
        The hyperplane arrangement corresponding to the 4-chain graph. The horizontal and vertical axes are
        the coordinate axes of the variables $S_{12}$ and $S_{34}$, respectively.
        The solid colored lines are the vanishing loci of the indicated energy factors $S_\tau$ while the dashed curve is the vanishing locus of the adjoint polynomial in~\eqref{eq:adjoint}.
        The horizontal and vertical dotted lines are the vanishing loci of~\eqref{eq:adjfact1} and~\eqref{eq:adjfact2},
        respectively.
        Left: arrangement for generic values of the parameters $(S_2,S_3,S_{1234})$.
        Right: degeneration of the arrangement obtained by shrinking an external bounded region (specifically, the
        triangle on the right) by setting $S_{23} + S_{34} - S_{234} = S_3=0$.
        Note how the adjoint polynomial factors such that one factor lies on top of the horizontal dotted line.
        Similarly at $S_{12} + S_{23} - S_{123} = S_2 = 0$ the adjoint polynomial factors such that
        one factor lies on top of the vertical dotted line.
    }
    \label{fig:4chainAdj}
\end{figure}

The numerator of a canonical function is called the \emph{adjoint polynomial} of the associated polytope.
For example, bringing the expression~\eqref{eq:4chainrecursive} for the 4-chain stripped wavefunction $\swc_{4\ch}$ onto a common denominator gives
\begin{align}
\label{eq:adjoint}
\begin{aligned}
    \swc_{4\ch} &=
    \frac{\text{adj}}{S_{12} S_{23} S_{34} S_{123} S_{234} }
    \,,
    \\
     \text{adj} &= S_{12} \left(S_{123} \left(S_{23}+S_{34}\right)+S_{234} S_{34}\right)+S_{23} S_{234} \left(S_{123}+S_{34}\right).
\end{aligned}\end{align}
The corresponding cosmological graph associahedron $\tilde{\A}_{4\ch}$ is the interior of the region bounded by
the hyperplanes $S_{12} = S_{23} = S_{34} = S_{123} = S_{234} = 0$ corresponding to the denominator
factors in~\eqref{eq:adjoint}.
For generic positive values of the parameters $(S_2, S_3, S_{1234})$, this region is the pentagon
in $(S_{12},S_{34})$-space depicted in figure~\ref{fig:4chainAdj}.
The vanishing locus of the adjoint polynomial, shown as a dashed line, passes through all points where the hyperplanes intersect outside of the pentagon; this property ensures that the canonical form has zero residue at these points.

According to the analysis of section~\ref{sec:cosmologicalzeros}, we know that $\swc_{4\ch}$ has (by factorizing the graph at the 2nd or 3rd site) factorization zeros at
\begin{align}
    S_{2} = S_{1234}+S_3-S_{12} = 0
    \qquad
    \text{and}
    \qquad
    S_{3}
    = S_{1234}+S_2-S_{34}
    = 0\,,
    \label{eq:sec4expected}
\end{align}
and a wavefunction zero at
\begin{align}
\label{eq:sec4wfz}
S_{123} - S_{12} + S_{34} = S_{123} + S_{234} = 0\,.
\end{align}
Here we want to think of these as conditions on the variables $(S_{12}, S_{34})$ for fixed values of the parameters $(S_2, S_3, S_{1234})$; therefore we use~\eqref{eq:linRel} to reexpress~\eqref{eq:sec4wfz} as
\begin{align}
\label{eq:sec4wfz2}
\begin{aligned}
    S_{123} + S_{234}
    &= (S_{1234}+S_2-S_{34})
    + (S_{1234}+S_3-S_{12}) = 0\,,
    \\
    S_{123}-S_{12}+S_{34}
    &= S_{1234}+S_3-S_{12} = 0
    \,.
\end{aligned}\end{align}

In general the adjoint polynomial is a high degree polynomial whose roots may be difficult to analyze, while
the wavefunction and factorization zeros are always linear conditions on the variables.
In order to see how it encodes the wavefunction and factorization zeros in our example,
let us take a limit in the parameters that shrinks
one of the external bounded regions in the hyperplane arrangement.  When this happens the vertices forming the region collide and become part of the polytope; such a degeneration of the hyperplane arrangement leads to a factorization of the adjoint polynomial.

For example, setting $S_3 = S_{23} + S_{34} - S_{234}= 0$ leads to the configuration in the right of figure~\ref{fig:4chainAdj}, where the triangle to the right of the pentagon has shrunk.
In this limit,
the adjoint polynomial factors as
\begin{align}
\label{eq:adjfact1}
    \text{adj}\vert_{S_{3}=0}
    = S_{234} S_{1234}
    \left(S_{1234}+S_2-S_{34}\right)
    \,,
\end{align}
where we have used~\eqref{eq:linRel}.
The first factor above cancels the $S_{234}$ pole in the denominator of $\swc_{4\ch}$ while the vanishing locus of the last factor corresponds to the horizontal dotted line in figure~\ref{fig:4chainAdj}.
This provides a geometric interpretation for the second factorization
zero listed in~\eqref{eq:sec4expected}.

Similarly setting $S_2=S_{12} + S_{23} - S_{123} = 0$ shrinks the triangle above the pentagon in figure~\ref{fig:4chainAdj} and the adjoint polynomial factors as
\begin{align}
\label{eq:adjfact2}
    \text{adj}\vert_{S_{2}=0}
    = S_{123} S_{1234}
    \left(S_{1234}+S_3-S_{12}\right)\,,
\end{align}
where we have again used~\eqref{eq:linRel}.
The first factor above cancels the $S_{123}$ pole in the denominator of $\swc_{4\ch}$ while the vanishing locus of the last factor corresponds to the vertical dashed line in the figure.
This provides a geometric interpretation for the first factorization
zero listed in~\eqref{eq:sec4expected}.

There is a third external bounded region -- the triangle to the upper right of the pentagon -- that can be shrunk by setting the total energy parameter $S_{1234}$ to zero.  In this limit the adjoint polynomial again factors, but one of the factors
cancels one of the poles in the stripped wavefunction:
\begin{align}
\swc_{4\ch}\rvert_{S_{1234} = 0}
= \frac{S_2 S_3 (S_2{+}S_3{-}S_{12}{-}S_{34})}{S_{12}( S_2{+}S_3{-}S_{12}{-}S_{34})S_{34}(S_3{-}S_{34})(S_2{-}S_{12}) } =
\frac{S_2 S_3}{S_{12}S_{34}(S_3{-}S_{34})(S_2{-}S_{12}) }\,,
\end{align}
so there is no novel zero associated with this limit; we do however see the two parametric zeros~\eqref{eq:parchain4} of $\swc_{4\ch}$ from this formula.

Finally we turn to the wavefunction zero~\eqref{eq:sec4wfz2}, and we see now that it
has a trivial explanation: it lies at the
intersection of the two factorization zeros listed in~\eqref{eq:sec4expected}!
For chain graphs it is always the case that the wavefunction zero lies at an intersection of
different factorization zeros, but this does not happen for other graph topologies.
This may be one way to see why chain graphs ``accidentally'' have wavefunction zeros that general graphs do not.

\section{Cosmological ABHY associahedra}
\label{sec:cosmoABHY}

In this section, we present a surprising connection between the (tree-level) scattering \emph{amplitudes} of a colored (i.e., matrix-valued) scalar field with a Tr($\phi^3$) interaction (which have full Poincar\'e symmetry) and the stripped cosmological wavefunctions (which lack time-translation invariance) associated to \emph{individual} chain graphs.
We begin in section~\ref{sec:ABHYreview} with a lightning review of the color-ordered amplitudes in Tr($\phi^3$) theory, the geometries that encodes them (the ABHY associahedra), and their hidden zeros discovered in~\cite{Arkani-Hamed:2023swr}.
In section~\ref{sec:wavefunctoamp}, we present an explicit map between the kinematic variables appearing in the $n$-chain stripped wavefunction and those of the $(n{+}1)$-point color-ordered amplitude in Tr($\phi^3$) theory before explaining how the cosmological zeros studied in this paper map to their amplitude counterparts.

\subsection{\texorpdfstring{A review of the hidden zeros of Tr($\phi^3$) theory}{A review of the hidden zeros of Tr(phi-cubed) theory}}
\label{sec:ABHYreview}

\begin{figure}[ht]
    \centering
    \includegraphics[scale=0.9]{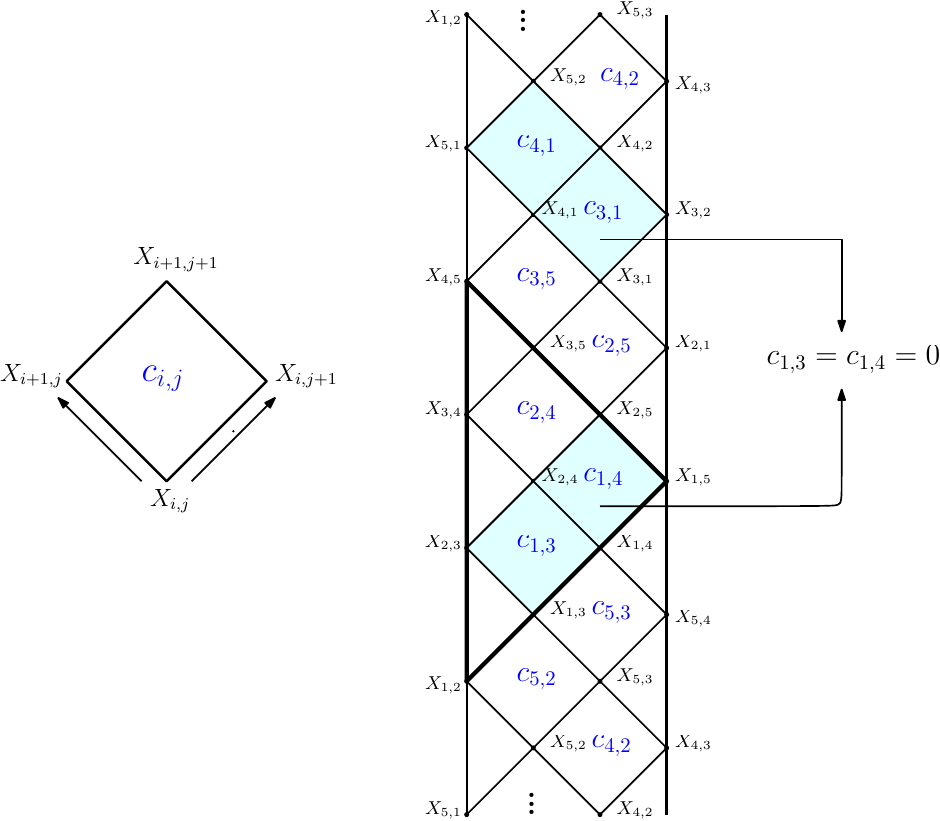}
    \caption{The 5-point kinematic mesh (right) with legend (left). The kinematic basis associated to the triangular region highlighted in bold is $\{X_{1,3}, X_{1,4}, c_{1,3}, c_{1,4}, c_{2,4}\}$. As an example of a zero of the amplitude $A_{5}^{\text{Tr}(\phi^3)}$, we highlight the case of $c_{1,3} = c_{1,4} =0$ associated to the maximal rectangle in blue.}
    \label{fig:5ptkinematicmeshABHY}
\end{figure}

\paragraph{Kinematic mesh.} Each $n$-point color-ordered amplitude in Tr($\phi^3$) theory, denoted by $A_{n}^{\text{Tr}(\phi^3)}$, is a function of the $n(n{-}1)/2$ Lorentz-invariant dot products of momenta $p_i\cdot p_j$. Thanks to momentum conservation only $n(n{-}3)/2$ of these are independent (in arbitrary spacetime dimension), which coincides with the number of \emph{planar kinematic variables} $X_{i,j}$\footnote{These should not be confused with the site energies $X_v$ of cosmological wavefunctions in~\eqref{eq:wfc}.} defined by
\begin{align}
    X_{i,j} = (p_i + \dots + p_{j-1})^2\,,
\end{align}
where we take the momenta to be on-shell so that $X_{i,i+1}=p_i^2 = 0$ and subscripts are always understood mod $n$. These planar variables comprise the complete set of propagators that appear in $A_{n}^{\text{Tr}(\phi^3)}$. We then define the \emph{non-planar variables} $c_{ij}\equiv -2p_i\cdot p_j$ for $|i{-}j|\ge2$, which can be expressed in terms of the $X_{i,j}$ via
\begin{align}
\label{eq:nonplanarABHY}
    c_{i,j}=X_{i,j}+X_{i{+}1,j{+}1}-X_{i,j{+}1}-X_{i{+}1,j}\,.
\end{align}
This identity motivates organizing both sets of variables in a~\emph{kinematic mesh} \cite{Arkani-Hamed:2017mur, Arkani-Hamed:2019vag}, where the planar variables label vertices and the non-planar variables are associated to specific diamonds, as seen in figure~\ref{fig:5ptkinematicmeshABHY} for the 5-point case.

\paragraph{ABHY associahedron.} A \emph{kinematic basis} is a choice of $n{-}3$ planar variables and $(n{-}2)(n{-}3)/2$ non-planar variables that are independent, which means that all of the $X_{i,j}$ can be expressed uniquely in terms of the variables in a basis. The former are treated as coordinates on an $(n{-}3)$-dimensional subspace of the $n(n{-}3)/2$-dimensional kinematic space, while the latter are thought of as parameters that specify the hypersurface.
The ABHY construction~\cite{Arkani-Hamed:2017mur} realizes the associahedron in this hyperplane as an $(n{-}3)$-dimensional polytope  whose canonical function equals the amplitude $A_{n}^{\text{Tr}(\phi^3)}$. For example at $n=5$ the associahedron is a pentagon (identical to the 4-chain graph associahedron from figure~\ref{fig:4sitegraphassockin}) and the associated canonical function is
\begin{equation}
    A_{5}^{\text{Tr}(\phi^3)} = \frac{1}{X_{1,3} X_{1,4}} + \frac{1}{X_{2,4} X_{2,5}} + \frac{1}{X_{1,3} X_{3,5}} + \frac{1}{X_{1,4} X_{2,4}} + \frac{1}{X_{2,5} X_{3,5}}\,.
    \label{eq:5ptabhyamplitude}
\end{equation}

\paragraph{Zeros.} The zeros of Tr$(\phi^3)$ amplitudes are reached by picking a maximal rectangular region in the kinematic mesh and setting to zero all of the $c_{i,j}$ inside it~\cite{Arkani-Hamed:2023swr}. In the 5-point example, one such zero is reached by setting $c_{1,3}=c_{1,4}=0$, which is associated to the maximal rectangle with top and bottom vertices $X_T=X_{2,5}$ and  $X_B=X_{1,3}$ respectively, as shown in figure~\ref{fig:5ptkinematicmeshABHY}.

In addition, there is a factorization property associated to a near-zero configuration which states that on the locus where all of the non-planar variables except one in a maximal triangle are zero, the amplitude factors into
\begin{align}
\label{eq:ABHYsplit}
    A_{n}^{\text{Tr} (\phi^3)} \rightarrow \left(\frac{1}{X_B} + \frac{1}{X_T}\right) \times A_{\text{down}}^{\text{Tr} (\phi^3)} \times A_{\text{up}}^{\text{Tr} (\phi^3)}~,
\end{align}
with appropriate kinematics for the up- and down- sub-amplitudes.
In the example shown in the figure, i.e.~the zero of $A_{5}^{\text{Tr} (\phi^3)}$ at $c_{1,3} = c_{1,4}=0$, relaxing the first condition leads to
\begin{align}
    A_5^{\text{Tr} (\phi^3)} \rvert_{c_{1,4} = 0}  = \frac{c_{1,3}}{X_{1,3} X_{2,5}} \times \left(\frac{1}{X_{1,4}} + \frac{1}{X_{3,5}}\right),
    \label{eq:ABHYsplittingex1}
\end{align}
while relaxing the second condition leads to
\begin{align}
    A_5^{\text{Tr} (\phi^3)} \rvert_{c_{1,3} = 0} \frac{c_{1,4}}{X_{1,3} X_{2,5}} \times \left(\frac{1}{X_{2,4}} + \frac{1}{X_{3,5}}\right).
    \label{eq:ABHYsplittingex2}
\end{align}
We refer the reader to \cite{Arkani-Hamed:2023swr} for a detailed analysis of these zeros and how they can be realized as flattening limits of the ABHY associahedron.

\subsection{From cosmological to amplitudes zeros}
\label{sec:wavefunctoamp}

It is already clear that the Tr($\phi^3$) story closely resembles that of the cosmological zeros we studied in section~\ref{sec:graphassoccosmo}.
We now present a concrete connection between the stripped wavefunctions for $n$-chains and the color-ordered scattering amplitudes in Tr$(\phi^3)$ theory (similar connections have been discussed recently in~\cite{Glew:2025otn, Glew:2025ugf}).
To start with we identify the planar variable $X_{i,j}$ of $A_{n+1}^{\text{Tr}(\phi^3)}$ with the energy
factor $S_\tau$ associated to the tube $\tau=\{i,i{+}1,\dots,j{-}1\}$:
\begin{align}
         S_{i \dots j{-}1} \equiv X_{i,j} \qquad |i-j| \ge 2\,,
\label{eq:wfctoamp}
\end{align}
with $X_{i,i+1} = 0$ as discussed above. Note that this identification does not determine the
energy parameters $S_i$ associated to 1-tubes, and in particular does not require that we must take them to vanish.
It is easy to see that the stripped wavefunction $\swc_{n\ch}$ of the $n$-chain is identical
to the $(n+1)$-point Tr($\phi^3$) scattering amplitude $A_{n+1}^{\text{Tr}(\phi^3)}$ under~\eqref{eq:wfctoamp}.
For example,
the five terms in~\eqref{eq:4chainrecursive} for $\swc_{4\ch}$ become identical to the five terms in the 5-point amplitude $A_{5}^{\text{Tr}(\phi^3)}$ shown in~\eqref{eq:5ptabhyamplitude}.

Next let us understand how the non-planar variables  $c_{i,j}$  defined in~\eqref{eq:nonplanarABHY} map to the parameters $\{ S_2, S_3, \ldots, S_{n-1}, S_{12\dots n}\}$ of the $n$-chain.
Using the map~\eqref{eq:wfctoamp} along with the linear relations of the $n$-chain obtained from~\eqref{eq:linRel}, we observe that:
\begin{enumerate}
    \item The $n{-}2$ parameters associated to energy 1-tubes of interior sites map according to
    \begin{equation}
        S_i = c_{i-1, i+1} \qquad \text{for } 2 \le i \le n-1\,.
    \label{eq:StoCcond1}
    \end{equation}
    \item A total of $(n{-}1)(n{-}4)/2$ non-planar variables  vanish identically on the support of~\eqref{eq:linRel}:
    \begin{align}
    c_{1,4} = c_{1,5} = \cdots = c_{1,n-1} &= 0\,, \nn \\
    c_{i,i+3} = c_{i,i+4} = \cdots = c_{i,n} &= 0 \qquad \qquad \text{for } 2 \le i \le n-3\,.
    \label{eq:StoCcond2}
    \end{align}
    \item The total energy factor $S_{1 \dots n}$ maps to
    \begin{equation}
        S_{12\dots n} = c_{1,n}\,.
    \label{eq:StoCcond3}
    \end{equation}
    \item The remaining $n{-}3$ non-planar variables map to linear combinations of the parameters and  energy factors $S_{\tau \in \mathsf{\tilde{T}}}$ of the $n$-chain graph; these can be worked out on a case-by-case basis.
\end{enumerate}
We call the image of the $n$-chain cosmological graph associahedron $\tilde{\A}_{n\ch}$ under this map the \emph{cosmological ABHY associahedron}.

Note that the map explicitly breaks the $\mathbb{Z}_{n+1}$ cyclic symmetry of the amplitude $A_{n{+}1}^{\text{Tr}(\phi^3)}$. Thus unlike for the ABHY associahedron, different cyclic permutations of the cosmological ABHY associahedron are not related to each other by different choices of kinematic basis.
The form of the map given above is tuned to the choice of coordinates provided by the ray-like triangulation of $A_{n{+}1}^{\text{Tr}(\phi^3)}$.
We refer the reader to~\cite{Arkani-Hamed:2023swr} for additional details; for our purpose it is enough to state that
elements of the kinematic basis associated to this triangulation are those that live inside a maximal triangular region in the kinematic mesh (see figure~\ref{fig:5ptkinematicmeshABHY}).

Another key difference between the ABHY associahedron and its cosmological counterpart arises from the vanishing non-planar variables~\eqref{eq:StoCcond2}. The vanishing of these parameters of the ABHY associahedron causes vertices to collide, resulting in a non-simple polytope as discussed in section~\ref{sec:AG}.

\begin{figure}[ht]
    \centering
    \includegraphics[scale=0.8]{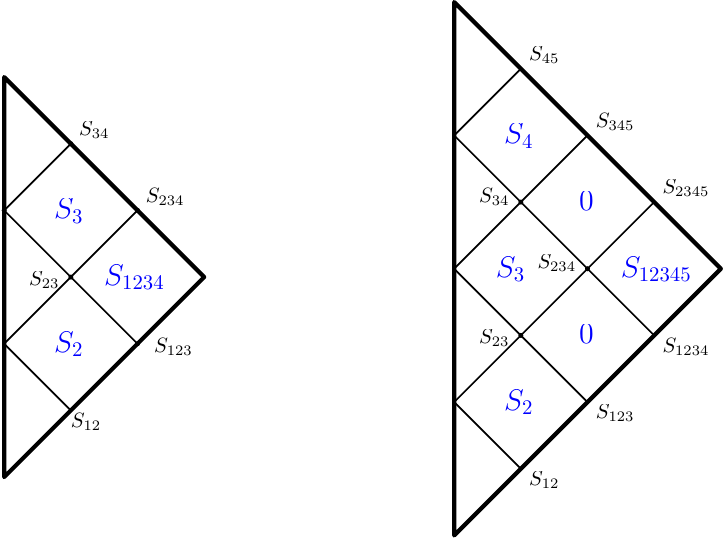}
    \caption{The ray-like subregion of the cosmological kinematic mesh for the 4- and 5-chain graphs.}
    \label{fig:cosmokinematicmesh}
\end{figure}

It is instructive to utilize our map between the kinematic variables $\{X_{i,j},c_{i,j}\}$ of Tr($\phi^3$) theory and
the energy factors $S_\tau$ of cosmological wavefunctions to translate the kinematic mesh associated to the
former into a \emph{cosmological kinematical mesh} for the stripped $n$-chain wavefunction $\swc_{n\ch}$.
Figure~\ref{fig:cosmokinematicmesh} shows, for $n=4,5$, the portion of the cosmological kinematic mesh
corresponding to the ray-like triangulation subregion of the full mesh.

Given this surprising connection between stripped wavefunctions for $n$-chains and $(n+1)$-point amplitudes in
Tr($\phi^3$) theory, it is natural to ask whether the cosmological zeros of the former, that we have
studied in section~\ref{sec:cosmologicalzeros}, map naturally to the zeros of the latter that were
characterized in~\cite{Arkani-Hamed:2023swr}.  We now illustrate how this works using the 4-chain as an example.

First we consider the parametric zeros~\eqref{eq:parChain}, which map to
\begin{align}
    S_2 = S_{1234} = 0 ~\rightarrow~ c_{1,3} = c_{1,4} = 0\,,  \label{eq:4chainparamzero1} \\
    S_3 = S_{1234} = 0 ~\rightarrow~ c_{1,4} = c_{2,4} = 0\,,
    \label{eq:4chainparamzero2}
\end{align}
reproducing two of the zeros of the amplitude $A_{5}^{\text{Tr}(\phi^3)}$. Note that like
for the $5$-point amplitude, the parametric zeros of the $4$-chain are also reached by setting all of
the parameters inside a maximal rectangle of the cosmological kinematic mesh to zero, shown in this
example in the left of figure~\ref{fig:cosmokinematicmesh}.
This is a generic feature -- setting the parameters inside a maximal rectangle in the ray-like subregion of the cosmological kinematic mesh for the $n$-chain enumerates all parametric zeros~\eqref{eq:parChain} for the stripped wavefunction $\swc_{n\ch}$, and these zeros correspond to a subset of the zeros of the $(n+1)$-point amplitude $A_{n+1}^{\text{Tr}(\phi^3)}$. Figure~\ref{fig:cosmokinematicmesh} (right) demonstrates this feature for $n=5$.

To access the other zeros of $A_{n+1}^{\text{Tr}(\phi^3)}$, one needs to go beyond the chosen ray-like subregion of the cosmological kinematic mesh. As discussed earlier, this does not correspond to a simple change of kinematic basis for the cosmological ABHY associahedron. The two factorization zeros for the 4-chain map to two more zeros of the 5-point amplitude according to
\begin{align}
    S_3 = S_{12} + S_{23} = 0 ~\rightarrow~ c_{2,4} = c_{2,5} = 0\,,  \\
    S_2 = S_{23} + S_{34} = 0 ~\rightarrow~ c_{1,3} = c_{3,5} = 0\,,
\end{align}
while the sole wavefunction zero~\eqref{eq:nchain_wfzero} gives the final zero of $A_5^{\text{Tr} (\phi^3)}$:
\begin{align}
    S_{123} - S_{12} + S_{34} = S_{123} + S_{234} = 0 ~\rightarrow~ c_{2,5} = c_{3,5} = 0\,.
\end{align}
Thus, we see that the zeros of the stripped wavefunction $\swc_{4\ch}$ associated to the single 4-chain graph non-trivially map to all five zeros of the 5-point amplitude $A_5^{\text{Tr} (\phi^3)}$,
which is given by a sum of contributions from 5 distinct graphs.

Again, this is a generic feature that extends to all $n$. In addition to the zeros, the near-zero factorization associated to Tr$(\phi^3)$ amplitudes is also reproduced by the stripped wavefunction: one can readily check that the factorization associated to the near-zero loci of the stripped wavefunction, described in section~\ref{sec:cosmologicalzeros}, reproduces the factorization property of the Tr$(\phi^3)$ amplitudes given in~\eqref{eq:ABHYsplit}. The only subtlety in checking this is that for generic $n$-chains, the near-zero factorization proceeds straightforwardly except for the fact that one needs to account for the $c_{i,j}$ variables that vanish at the outset~\eqref{eq:StoCcond2}. These coincide with the non-simple vertices of the cosmological ABHY associahedron. Given a maximal rectangle, one cannot turn these vanishing $c_{i,j}$ back on; however, the other non-trivial $c_{i,j}$ variables can be turned on in the usual manner to observe the near-zero factorization described above.

\section{Discussion and outlook}
\label{sec:outlook}

In this paper, we have classified the linear zero conditions for the flat-space (stripped) wavefunction coefficients (associated to a single Feynman graph $\G$) into parametric, wavefunction and factorization zeros.
The wavefunction zeros (section~\ref{sec:wav0}) are exclusive to graphs of chain topology and cause both the wavefunction and its stripped counterpart to exhibit a universal splitting behavior on its near-zero locus.
The factorization zeros (section~\ref{sec:fac0}) rely on first factorizing an arbitrary graph $\G$ solely into its constituent chain subgraphs and then localizing on to the wavefunction zeros of all such chain subgraphs.
Finally, the parametric zeros (section~\ref{sec:par0}) of a graph $\G$ are seen to arise as flattening limits of the associated cosmological graph associahedron $\tilde{\A}_{\G}$ -- a non-simple polytope obtained from a certain cosmological limit of the graph associahedron $\A_\G$.
While the wavefunction and factorization zeros are not discoverable as flattening limits of $\tilde{\A}_{\G}$, they still admit geometric interpretations: new zero conditions on the coordinates are discovered by degenerating the hyperplane arrangement so that the adjoint polynomial (numerator of the wavefunction) factors into linear pieces.
Finally, we see a surprising connection between the stripped wavefunction coefficients of chain graphs to scattering amplitudes in Tr$(\phi^3)$ theory. This connection remarkably allows us to use the zeros for the wavefunction associated to a single graph to find the zeros of amplitudes in Tr$(\phi^3)$ theory, which involve a sum over many Feynman graphs.

We note that while we have focused on an uncolored scalar theory with a $\phi^3$ interaction, the analysis of this paper generalizes to arbitrary polynomial $\phi^p$ interactions in a straightforward manner. Indeed the zero loci of specific graphs remain unchanged upon changing the valence of the scalar self-interaction. On the other hand, the definitions of the energy factors in terms of which these zero loci are expressed will have to be changed accordingly.

While the central objects of interest in this paper have been the flat-space wavefunction coefficients, it is important to note that they serve as universal integrands for their counterparts in a generic power-law FRW spacetime with scale factor $a(\eta)=\eta^{-1-\varepsilon}$ and cubic interactions%
\footnote{The integration kernel depends on the valency of the interaction.}
\begin{align}
    \wc_{n,\text{FRW}}^{(\ell)} = \int_0^\infty \ \left(\prod_v \text{d}x_v \ x_v^{\varepsilon}  \right) \wc_{n}^{(\ell)} (x_v + X_v, Y_e)~.
    \label{eq:FRWwavefunc}
\end{align}
Here, the site energies of the flat wavefunction (defined in \eqref{eq:wfc}) are shifted by $x_v$ and then integrated over the kernel $ x_1^{\varepsilon}\cdots x_n^{\varepsilon}$.
Consequently, the zeros of the universal integrand $\wc_{n}^{(\ell)}$ studied here naturally carry over to wavefunction coefficients in power-law FRW cosmologies $\wc_{n,\text{FRW}}^{(\ell)}$.

Also, while our analysis is centered around the flat-space wavefunction corresponding to a single Feynman graph $\G$, it would be natural to perform the same analysis of the zeros for the Tr$(\phi^3)$ cosmological wavefunction that involves a sum over Feynman diagrams~\cite{Arkani-Hamed:2024jbp}.
Just as the parametric zeros of $\G$ were discovered from flattening limits of the (non-simple) cosmological graph associahedron as discussed in section~\ref{sec:graphassoccosmo}, it would be interesting to study if the zeros of the Tr$(\phi^3)$ wavefunction arise from flattening the associated geometry -- the cosmohedron~\cite{Arkani-Hamed:2024jbp}. 

Unlike the parametric zeros and even the amplitude zeros, the wavefunction and factorization zeros are on a different footing as they are not directly accessible via flattening limits of the cosmological graph associahedron $\tilde{\A}_{\G}$. It would therefore be interesting to understand its geometrical origins that complements or extends the notion of flattening the associated polytope. Despite the vanishing of the canonical function of $\tilde{\A}_{\G}$, the polytope does not flatten on the loci of such zeros. It remains an open question whether these non-flattening limits admit a natural interpretation within the structure of the graph associahedron.

In section~\ref{sec:cosmoABHY}, we discussed that the graph associahedra for $n$-chains are simply the classical associahedra, which is of cluster type $A_{n-2}$, while the $n$-gons at 1-loop are of type $B_{n-1}$. 
Cluster string integrals~\cite{Arkani-Hamed:2019mrd} exist for such associahedra, paving the way for stringy formulations of wavefunction coefficients associated with single graphs in cosmology -- an active area of ongoing research.

From a more physical perspective, it is important to understand the physical origin of the zeros for the cosmological wavefunction studied in this paper. Recasting the conditions of the $n$-chain wavefunction zero  \eqref{eq:nchain_wfzero} explicitly in terms of its energy variables
\begin{align}
    X_i=0\ \text{ for }\ 3\le i\le n-2\,, && Y_1 = -X_1 - 2X_2\,, && Y_{n{-}1} = -2 X_{n{-}1}-X_n\,,
    \label{eq:wavefunczeroenergyvariables}
\end{align}
we observe that the associated kinematics exhibit multi-soft limits, i.e the energy of the external leg at the $i$-th site vanishes:  $X_i = |\vec{k}_i| = 0$ for $3 \le i \le n-2$. It would be interesting to see if the kinematics of such zeros could be exploited to better understand the soft behavior of the wavefunction coefficients. Also at the level of flat-space amplitudes, the existence of hidden zeros are linked to the color-kinematics duality and can be derived from the Bern-Carrasco-Johansson (BCJ) relations~\cite{Bartsch:2024amu}. To that end, understanding the symmetry arguments behind the cosmological zeros is an important direction we leave for future study.

Finally, it would also be interesting to see if the flat-space (stripped) wavefunction can be uniquely fixed from its zeros, as was recently shown to be true for amplitudes in Tr$(\phi^3)$ theory~\cite{Rodina:2024yfc}.

\acknowledgments

We are grateful to Nima Arkani-Hamed, Carolina Figueiredo, David Gross, Austin Joyce, Hayden Lee and Tomasz \L ukowski for helpful discussions. This work was supported in part by the US Department of Energy under contract DE-SC0010010 Task F (SD, AP, MS, AV), by Simons Investigator Award \#376208 (SP, AV), and by the National Science Foundation grant NSF PHY-2309135 to the Kavli Institute for Theoretical Physics (KITP) (SD, MS, AV).

\bibliographystyle{JHEP}
\bibliography{refs.bib}

\end{document}